\documentclass[twocolumn]{aastex62}

\usepackage{savesym}
\savesymbol{tablenum}
\usepackage[T1]{fontenc}
\usepackage{inputenc}
\usepackage[range-units = brackets,tophrase={-},seperr,repeatunits=false]{siunitx}
\restoresymbol{SIX}{tablenum}

\graphicspath{{./}}

\accepted{July 24, 2019}

\submitjournal{ApJ}

\shorttitle{A\,1758}
\shortauthors{Schellenberger et al.}

\begin{document}
	
	\title{Forming one of the most massive objects in the Universe: The quadruple merger in Abell 1758}

	\author[0000-0002-4962-0740]{G. Schellenberger}
	\correspondingauthor{Gerrit Schellenberger}
	\affil{Harvard-Smithsonian Center for Astrophysics, 60 Garden Street, Cambridge, MA 02138, USA}
	\email{gerrit.schellenberger@cfa.harvard.edu}

	\author{L. David}
	\affiliation{Harvard-Smithsonian Center for Astrophysics, 60 Garden Street, Cambridge, MA 02138, USA}

	\author[0000-0002-5671-6900]{E. O'Sullivan}
	\affiliation{Harvard-Smithsonian Center for Astrophysics, 60 Garden Street, Cambridge, MA 02138, USA}
	
	\author{J. M. Vrtilek}
	\affiliation{Harvard-Smithsonian Center for Astrophysics, 60 Garden Street, Cambridge, MA 02138, USA}
	
	\author{C. P. Haines}
	\affiliation{School of Physics and Astronomy, University of Birmingham, Edgbaston, Birmingham, B15 2TT, UK}		
	\affiliation{Instituto de Astronom\'{i}a y Ciencias Planetarias de Atacama, Universidad de Atacama, Copayapu 485, Copiap\'{o}, Chile}
		
	\begin{abstract}
		Abell\,1758 is a system of two galaxy clusters, a more massive, northern cluster and a southern cluster. Both parts are undergoing major merger events at different stages. Although the mass of the merger constituents provides enough energy to produce visible shock fronts in the X-ray, none have been found to date.
		We present detailed temperature and abundance maps based on \textit{Chandra} ACIS data, and identify several candidates for shocks and cold fronts from a smoothed gradient map of the surface brightness. 
		One candidate can be confirmed as the missing shock front in the northern cluster through X-ray spectroscopy.
		Non-thermal radio emission observed with the GMRT confirms the presence of radio halos in the northern and southern clusters, and shows evidence for a relic in the periphery of the southern cluster.
		We do not find evidence for shocked gas between A\,1758\,N and A\,1758\,S.
		
	\end{abstract}
	
	\keywords{galaxies: clusters: individual (A1758) -- X-rays: galaxies: clusters -- radio continuum: general -- galaxies: clusters: intracluster medium -- galaxies: interactions}

	\section{Introduction}\label{sec:intro}
	\subsection{Galaxy clusters in filaments and mergers}
	Thermal X-ray emission from the intracluster medium (ICM), i.e., gas accreted through the dark matter potential and heated to keV temperatures, provides information on the dynamical history of the cluster in the form of densities, temperatures, and heavy element abundances. Comparing maps of the thermodynamic structure of a cluster with hydrodynamic simulations yields important conclusions about merger geometry. This has been shown, e.g., in \cite{2009MNRAS.399.1307M}, for a comparison of relaxed and unrelaxed clusters.
	Information on temperature and density of the gas can be used to derive the total cluster mass, assuming the ICM is in mechanical equilibrium with the gravitational potential.
	
	Galaxy cluster systems with more than one component (e.g., cluster pairs) are of particular interest since they likely mark cosmic filaments, which are otherwise difficult to confirm observationally, especially in X-rays (expected gas temperatures $10^5-10^7\,\si{K}$). It is important to determine the amount of baryonic matter within filaments and the link between galaxy clusters and the cosmic web structure. To date, only few such galaxy cluster pairs with joining filamentary structure have been observed (e.g., \citealp{2005ApJ...619..161G,2010A&A...517A..94D,Bonjean2018-dd}).
	Galaxy properties (e.g., star formation, dynamical structure) should also be included within this picture: are galaxies in the filaments different from field galaxies? \cite{2016MNRAS.455..127M} suggest that the quenching of star formation already starts in the filaments, but the cluster environment shows even lower star formation rates than filaments (\citealp{2008ApJ...672L...9F}).
	Simulations of cluster pairs indicate that for separations of less than $\SI{5}{Mpc}$ there is always a connecting filamentary structure (\citealp{2005MNRAS.359..272C}).

	In contrast to the constant accretion of matter through filaments, violent cluster merging events leave very clear imprints in the ICM: shocks as the result of a supersonic collision with other clusters or groups can be detected as pressure discontinuities in X-ray density and temperature maps. Many examples have been reported in the past (e.g., \citealp{2007PhR...443....1M}). The interaction leads to the production of relativistic particles, which are detected through non-thermal synchrotron emission in radio observations. 
	Shocks have also been associated with diffuse radio emission in galaxy clusters, so called radio relics. Located in the cluster outskirts, with	elongated shapes and polarized radio emission with steep spectra, radio relics have been identified with remnants of energetic shock waves (\citealp{1998A&A...332..395E}).
	Other sources of diffuse radio emission in galaxy clusters are radio halos, which span up to megaparsec scales. They have been associated with dynamically disturbed clusters exhibiting cluster-wide particle acceleration processes (\citealp{2012A&ARv..20...54F}). However, not all dynamically disturbed clusters show large radio halos. 
	
	\subsection{Abell\,1758}
	\label{ch:A1758}
	Abell\,1758 is a rich galaxy cluster system at redshift $\num{0.279}$ that has been well studied at many wavelengths. It was discovered by \cite{1958ApJS....3..211A} as a richness class 3 cluster, and later identified as a complex system by \cite{1998MNRAS.301..328R} using data from ROSAT. 
	
	\cite{2004ApJ...613..831D} analyzed \textit{Chandra} and \textit{XMM-Newton} data confirming that A\,1758  is divided into a northern (N) and a southern (S) cluster with about $\SI{2}{Mpc}$ separation. Each of the two clusters is undergoing a merger: A\,1758\,N is hotter ($\SI{7}{keV}$) and likely to be in a post-core-passage stage, with the merger occurring almost in the plane of the sky. 
	A\,1758\,S (about $\SI{5}{keV}$) is in an earlier stage, where the two components are not clearly separated and the overall shape is not as elliptical as the northern cluster. In this paper we refer to Abell\,1758 as a system consisting of two clusters, not as a single cluster.
	
	The northern cluster is clearly the more dominant in terms of mass and X-ray luminosity (\citealp{2004ApJ...613..831D}), but itself consists of two subclusters (A\,1758\,NW and A\,1758\,NE), of which the north-western subcluster is the more massive and brighter, and appears to be the main component. The northern merger between NW and NE may be characterized by a large impact parameter, although this might be in contradiction with elongated regions of higher metallicity created by ram pressure stripping. 
	\cite{2011A&A...529A..38D} created temperature and abundance maps of A\,1758\,N and S from \textit{XMM-Newton} data, detecting two regions of enhanced metallicity in the outskirts of A\,1758\,N. The authors concluded that the galactic winds are an effective way to produce the high metallicity outer regions.
	
	Significant results for A\,1758 have also been reported in the radio regime: \cite{1985AJ.....90..927O} found a central narrow angle tail radio galaxy with an inferred direction of motion toward the center of A\,1758\,N in a high resolution VLA observation. 
	With low frequency data from the Giant Metrewave Radio Telescope (GMRT), \cite{2013A&A...551A..24V} could classify the diffuse radio source in A\,1758\,N as a giant radio halo with a spectral index\footnote{defined by $f_\nu \propto \nu^{\alpha}$, where $f_\nu$ is the spectral flux density and $\nu$ is the frequency} $\alpha = -1.3$ and an elongation along the merger axis, but could not confirm the radio relics: the emission might be part of the complicated structured radio halo. 
	Very recently \cite{Botteon2018-yp} confirmed the northern halo at three frequencies with LOFAR, GMRT, and JVLA ($144$, $\SI{325}{MHz}$ and $\SI{1.4}{GHz}$ respectively) and measured a spectral index of $-1.2$. The authors also report the detection of a radio halo for A\,1758\,S and a relic east of the southern cluster. 
		
	Several weak lensing studies have confirmed the high mass of this extraordinarily-structured cluster (\citealp{2002ApJS..139..313D,2008PASJ...60..345O,2017MNRAS.466.2614M}). Main findings include the NW subcluster being the main structure in the northern cluster, a weak lensing mass for A\,1758\,S, and an age after merger of the northern cluster of 270\,Myr.  This is in agreement with the simulations of a A\,1758\,N-like system presented in \cite{2015MNRAS.451.3309M}. This simulation suggest that the two parts of the northern cluster approached with an infalling velocity of $\SI{1500}{km\s^{-1}}$ and an impact parameter of $\SI{250}{kpc}$. The current relative velocity between the NW and NE clumps is predicted to be $\SI{380}{km\,s^{-1}}$, and the turnover point is almost reached. 
	Recently, \cite{Haines2018-re} confirmed an infalling subgroup with a mass $\SI{4.3e13}{M_\odot}$ of onto the northern cluster. We summarize all reference masses for A\,1758 in Table \ref{tab:masses}.
	
	Spectroscopic redshift measurements of member galaxies have been used in the past to understand the dynamics of this merging cluster: \cite{Boschin2012-vd} focused their galaxy dynamics analysis on A\,1758\,N, where they do not detect a bimodal distribution in the velocities, but find small groups of galaxies in the vicinity of the cluster.
	
	Despite the large number of studies at multiple wavelengths, there are some outstanding, but still unsolved, issues connected with this system, which we will address in this paper:
	are the two clusters, A\,1758\,N and S building a bound system? If so, can one detect large scale filaments by looking at the galaxy populations around and between the northern and southern cluster? Detecting residual X-ray emission between those will indicate the state of the entire system, if it is gravitationally bound.
	Although we are looking at a very massive merger, and there seems to be some extended radio emission present, no clear shock front has been detected up to now. Detailed temperature and abundance maps, as well as new techniques for finding surface brightness discontinuities, will help to detect shocks in X-rays. Since unpublished radio data ($\SI{608}{MHz}$ GMRT) are available, the emission from the giant radio halo, and from possible relics, can be investigated in more detail.
	The merging scenario of the southern cluster is still unclear due to the X-ray morphology. The deeper \textit{Chandra} data will help to find possible shock fronts.
	The radio analysis can confirm the halos and the relic at $\SI{608}{MHz}$, and provide an independent measurement.
	Finally, the wide range of studies in different wavelengths enables a detailed mass comparison to understand the geometry along the line of sight.
	
    In section \ref{ch:data} we outline our data analysis strategy. In section \ref{ch:results} we present the results separately for the X-ray data (investigating thermodynamic and chemical structure, and the search for shock fronts), the radio data (detecting the radio halos and relic in the system), and optical/spectroscopic data (revealing the dynamical structure and local galaxy overdensities). 
    We discuss the impact of the radio halos and the ICM structure in section \ref{ch:disc}, and show the final conclusions in section \ref{ch:conclusion}.

	Throughout this paper we assume a flat $\Lambda$CDM cosmology with the following parameters:\\$\Omega_{\rm m} = \num{0.27}$, $\Omega_\Lambda = \num{0.73}$, $H_0 =  h \cdot \SI{100}{km~s^{-1}~Mpc^{-1}}$ with $h = \num{0.7}$, and uncertainties that are stated at the 68\% confidence level unless otherwise indicated.
	\begin{deluxetable}{cccc}
		\tablecaption{Masses for A\,1758 \label{tab:masses}}
		\tablehead{
			\colhead{cluster} & \colhead{method} & \colhead{$\frac{M_{200}}{\SI{e15}{M_\odot}} $} & \colhead{reference}
		}
		\startdata
		N+S & Xray M-YX  &  2.6 & {\footnotesize\cite{2004ApJ...613..831D}}\\
		N+S & Xray hydro &  3.4 & this work \\
		N+S & SZ/Planck  & 1.2 & {\footnotesize\cite{2015arXiv150201598P}}\\
		N+S & SZ/AMI & 1.54 & {\footnotesize\cite{2012MNRAS.425..162A}}\\
		N+S & dynamic  & 2.5 & {\footnotesize\cite{2013ApJ...775..126H}}\\
		N+S & WL  & 3.0 & {\footnotesize\cite{2002ApJS..139..313D}}\\
		N+S & WL & 1.8 & {\footnotesize\cite{2017MNRAS.466.2614M}}\\
		N & Xray M-T  & 1.6 & {\footnotesize\cite{2004ApJ...613..831D}} \\
		N & Xray  & 1.8& {\footnotesize\cite{2014MNRAS.443.2342M}}\\
		N & Xray M-YX  &  0.74 & {\footnotesize\cite{2017ApJ...846...51L}}$\mathsection$ $\dagger$\\
		N & SZ/AMI  & 0.59 & {\footnotesize\cite{2012MNRAS.425..162A}} \\
		N & Xray/isoth. & 1.7 & {\footnotesize\cite{Zhang2008-fl}}$\dagger$\\ 
		N & Xray hydro & 2.3 & this work \\
		N & WL  & 2.9 & {\footnotesize\cite{2006ApJ...653..954D}}$\star$\\
		N & WL  & 4.5 & {\footnotesize\cite{2007ApJ...667...26P}}$\star$\\
		N & WL  & 0.70& {\footnotesize\cite{2008PASJ...60..345O}}$\star$\\
		N & virial  & 2.8 & {\footnotesize\cite{Boschin2012-vd}}\\
		N & WL  & 2.2 & {\footnotesize\cite{2012ApJ...744...94R}}\\
		N & WL  & 2.2 & {\footnotesize \cite{2014MNRAS.439...48A}}$\star$\\
		N & WL  & 1.6 & {\footnotesize\cite{2015MNRAS.449..685H}}\\
		N & WL  & 1.3 & {\footnotesize\cite{2017MNRAS.466.2614M}}\\
		N & WL  & 0.9 & {\footnotesize\cite{2016MNRAS.461.3794O}}\\
		NW & Xray hydro & 1.04 & this work \\
		NW & virial  & 1.4 & {\footnotesize\cite{Boschin2012-vd}}\\
		NW & WL  & 0.79 & {\footnotesize\cite{2017MNRAS.466.2614M}}\\
		NE & Xray hydro & 1.24 & this work \\
		NE & virial  & 0.7 & {\footnotesize\cite{Boschin2012-vd}}\\
		NE & WL  & 0.55 & {\footnotesize\cite{2017MNRAS.466.2614M}}\\
		S & Xray M-T  & 1.0 & {\footnotesize\cite{2004ApJ...613..831D}} \\
		S & Xray M-YX  &  0.95 & {\footnotesize\cite{2017ApJ...846...51L}}$\mathsection$ $\dagger$\\
		S & Xray hydro & 1.06 & this work \\
		S & SZ/AMI  & 0.63 & {\footnotesize\cite{2012MNRAS.425..162A}}\\
		S & WL  & 0.14 &{\footnotesize\cite{2008PASJ...60..345O}}$\star$\\
		S & WL  & 0.50 & {\footnotesize\cite{2017MNRAS.466.2614M}} \\
		\enddata
		\tablecomments{$\star$: as reported by \cite{2015MNRAS.450.3633S}; $\dagger$: $M_{500}$ conversion from \cite{Suhada2011}; \mbox{$\mathsection$: private} communication}
	\end{deluxetable}

	\section{Data reduction and analysis}
	\label{ch:data}
	Our X-ray analysis is based on four \textit{Chandra} and one \textit{XMM-Newton} observation (Table \ref{tab:data}). As described in \cite{2004ApJ...613..831D} the ACIS observation 2213 is entirely flared and was only used for imaging, while for the spectral analysis $\SI{146}{ks}$ \textit{Chandra}/ACIS-I and $\SI{16}{ks}$ \textit{XMM}/EPIC are employed. For the detection of surface brightness discontinuities we rely only on the \textit{Chandra} data. 
	The radio study focuses on the GMRT $\SI{608}{MHz}$ data (see Table \ref{tab:radiodata}). 
	In the following we briefly describe our data reduction procedures.
	
	\subsection{Chandra}
	The available \textit{Chandra} observations (see Table \ref{tab:data}) sum up to a total exposure time of $\SI{220}{ks}$. 
	Our data reduction used the standard tasks provided with the CIAO software package version 4.8 with CALDB 4.7.1 and HEASoft 6.20, and we followed the procedure described in \cite{2017ApJ...845...84S}.
	The spectral background was subtracted by using the provided blank sky observations re-normalized by the count rate at high energies, $\SIrange{9.5}{12}{keV}$.
	For outer low surface brightness regions we use the stowed events files (recording the particle background) and simultaneously fitted the RASS spectra to them (as provided by the \verb|HEASARC| tool \verb|X-ray Background|\footnote{\url{http://heasarc.gsfc.nasa.gov/cgi-bin/Tools/xraybg/xraybg.pl}}).
	Point sources were detected via the \verb|wavdetect| task in each observation. Point sources with a significance of at least 3 and no association with any cluster core were cross-matched and subtracted/excluded from spectral and imaging analysis.
	
	\subsection{XMM-Newton}
	The $\SI{55}{ks}$ \textit{XMM-Newton} EPIC observation (OBSID 0142860201) was reduced by the default tasks \verb|emchain| and \verb|epchain| within the SAS software package (version 15.0.0).
	For MOS data we included all events with \verb|PATTERN <= 12| and used the \verb|#XMMEA_EM| flagging, while for PN data we included only \verb|PATTERN <=4| and \verb|FLAG==0|. The PN data were corrected for out-of-time events. The lightcurve cleaning was performed using the \verb|ESAS| tasks \verb|mos-filter| and \verb|pn-filter|, resulting in a 70\% reduction in useful exposure time.

	\begin{deluxetable*}{ccccc}
		\tablecaption{X-ray data for A\,1758 \label{tab:data}}
		\tablehead{
			\colhead{Instrument} & \colhead{Observation} & \colhead{Date} & \colhead{exposure} & \colhead{resolution}
		}
		\startdata
		\textit{Chandra}/ACIS-S & 2213 & 2001 August 28 & $(\SI{58}{ks})^\dagger$ & $\SI{0.5}{\arcsec}$ \\
		\textit{Chandra}/ACIS-I & 13997  & 2012 September 27 & $\SI{27}{ks}$ &$\SI{0.5}{\arcsec}$ \\
		\textit{Chandra}/ACIS-I & 15538 & 2012 September 28 & $\SI{93}{ks}$ & $\SI{0.5}{\arcsec}$\\
		\textit{Chandra}/ACIS-I & 15540 & 2012 October 9 & $\SI{26}{ks}$ & $\SI{0.5}{\arcsec}$\\
		\textit{XMM}/EPIC & 0142860201  & 2002 November 12 & $\SI{16}{ks}$ &  $\SI{10}{\arcsec}$ \\
		\enddata
		\tablecomments{Exposure time is after cleaning of flares, except for $^\dagger$: The entire observation 2213 was flared and not used for spectral analysis.}
	\end{deluxetable*}

	\subsection{GMRT}
	\begin{deluxetable*}{cccccccc}
		\tablecaption{Summary of the GMRT observation \label{tab:radiodata}}
		\tablehead{
			\colhead{Project} & \colhead{Frequency} & \colhead{Bandwidth} & \colhead{Date} & \colhead{Time} & \colhead{robust} & \colhead{FWHM} & \colhead{r.m.s} \\
			\colhead{}  &  \colhead{(MHz)} & \colhead{(MHz)} & \colhead{} & \colhead{(min)} & &\colhead{(\si{\arcsec} $\times$ \si{\arcsec})} & \colhead{(mJy beam$^{-1}$)} \\
		}
		\startdata
		23\_023  & 608 &   32 & 2007 August 31 & 240 &  0  & $11\times 5$  & 0.09 \\
		23\_023  & 608 &   32 & 2007 August 31 & 240 &  1.0 \& \textit{uv}-taper $\SI{20}{\arcsec}$   & $44\times39 $  & 0.2  \\
		\enddata
		\tablecomments{Column 1: project code. Columns 2--5: observing frequency, bandwidth, date and total time. Column 6: Weighting scheme (Briggs robust parameter and tapering). Column 7: full-width half maximum (FWHM) of the array. Column 8: image r.m.s. level ($1\sigma$).}
	\end{deluxetable*}
	
	The archival GMRT data were reduced using the automated \verb|SPAM| pipeline (\citealp{2009A&A...501.1185I}), which not only flags bad data and applies flux and phase calibrations to the target of interest, but also performs a self-calibration and peeling task to correct for time and sky-position dependent variations in the phase solutions, which is of particular importance at lower frequencies. For flux- and bandpass calibration the primary calibrator \verb|3C286| with the flux scale by \cite{2012MNRAS.423L..30S} was used. 
	A phase calibrator was observed several times between the target scans. The output of the pipeline was further processed with the Common Astronomy Software Applications (CASA; \citealp{2007ASPC..376..127M}) package in version 5.1.2. We applied further flagging of bad data with the \verb|aoflagger| (\citealp{2010ascl.soft10017O}) tool, and then performed several steps of phase self-calibration followed by an amplitude self-calibration step. 
	Through this technique we achieved a noise level of $\SI{90}{\mu Jy\,beam^{-1}}$. 
	Since we are interested in the extended emission, we first create an image of the compact sources in the field by using a lower uv-cut at $3\,k\lambda$ and using a briggs-baseline-weighting scheme with \verb|robust=-0.25|. 
	The corresponding model was subtracted from the calibrated visibilities. Our final image was created with a lower uv-cut at $50\lambda$, \verb|robust=1| and a Gaussian uv-tapering of $30\arcsecond$ (see also Tab \ref{tab:radiodata}). 
	All imaging was performed with the \verb|clean| task and a multi-frequency-synthesis (\verb|MFS|) algorithm with two Taylor-terms to model the frequency dependence (\verb|nterms=2|). 
	The widefield corrections (\verb|wproject|, see \citealp{2005ASPC..347...86C}) were applied with 593 projection planes. Sources were detected from an initial image with the \verb|PyBDSF| module (\citealp{2015ascl.soft02007M}) and added to a mask for further imaging. 
	As a last step, we corrected for the GMRT primary beam response\footnote{\url{http://www.ncra.tifr.res.in:8081/~ngk/primarybeam/beam.html}}.
	
	\subsection{Optical spectra}
	Galaxies in and around Abell\,1758 were targeted with the Hectospec instrument as part of the Arizona Cluster Redshift Survey (ACReS, \citealp{2013ApJ...775..126H}). Hectospec is a 300-fiber multi-object spectrograph with a 1-degree diameter circular field-of-view that is installed on the $\SI{6.5}{m}$ MMT telescope. Galaxies were selected using near-infrared photometry from UKIRT/WFCAM imaging that covers a $52^{\prime}{\times}52^{\prime}$ field, centred on the cluster. Probable cluster members were identified from their $J-K$ colour, which can be used as a cheap redshift proxy that is also independent of star-formation history \cite{2009MNRAS.396.1297H}, and fibers assigned preferentially to galaxies brighter than K=17.1 ($K^{*}+1.5$ at $z=0.28$), with the objective of obtaining stellar-mass limited samples. Six Hectospec configurations were observed, and reduced using HSRED, enabling redshifts for 1490 galaxies to be measured.

    Abell\,1758 was also targeted with Hectospec by \cite{2013ApJ...767...15R} as part of the Hectospec Cluster Survey (HeCS), obtaining redshifts for 424 galaxies selected to lie on the cluster red sequence (g-r vs $M_{r}$) across the same degree-wide field. Of the 289 galaxies observed by both surveys, 287 had measured redshifts that agreed within $\SI{300}{km\,s^{-1}}$ (>99\% success rate), enabling us to confirm that both surveys have typical redshift uncertainties of $\SI{50}{km\,s^{-1}}$ with no discernible systematic offset.
    
    \cite{2017MNRAS.466.2614M} observed A\,1758 with the Gemini Multi-Object Spectrograph (GMOS) mounted on the Gemini North telescope, placing four slit masks along a SE-NW line through A\,1758\,N, and a fifth covering the core of A\,1758\,S. Their final catalogue contains 203 redshift measurements for A\,1758\,N and 35 for the A\,1758\,S field. Of the 119 targets in common between \cite{2017MNRAS.466.2614M} and ACReS, 117 had measured redshifts that agreed within $\SI{500}{km\,s^{-1}}$ (>98\% success rate). 
    
\begin{figure}
        \begin{center}
		    \includegraphics[width=0.4\textwidth]{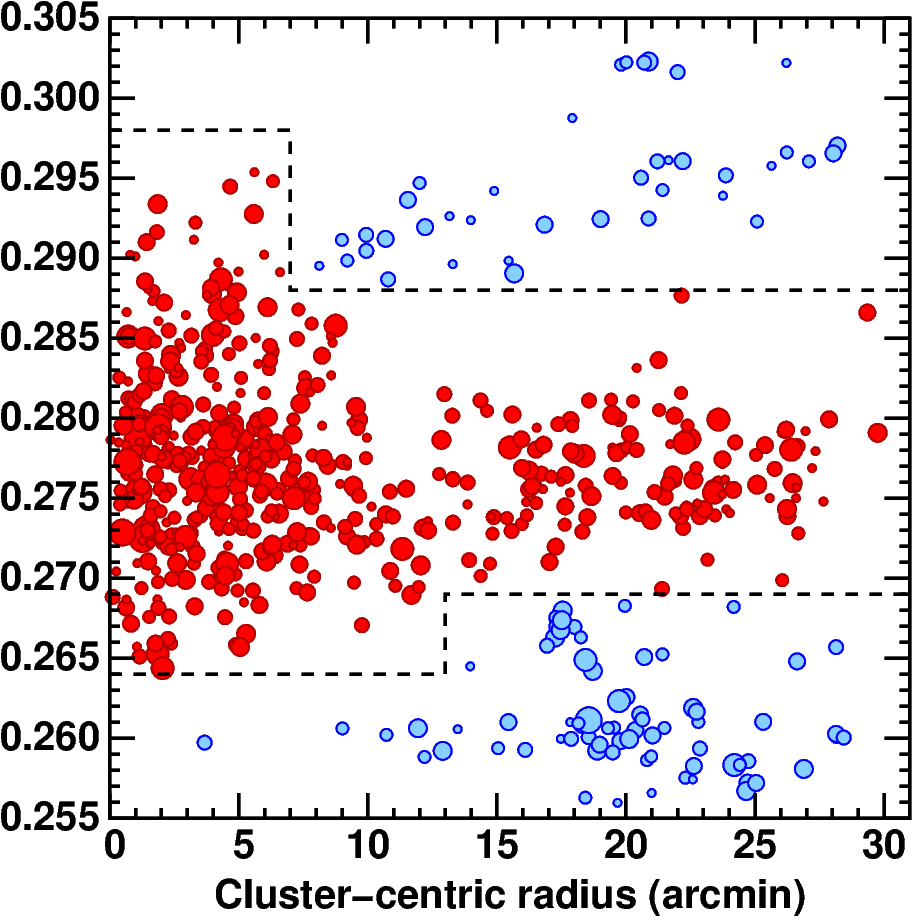}
		    \caption{Caustic diagram showing the galaxy redshifts versus distances to the nearest cluster center (north or south). Red points are identified as members of A\,1758\,N/S, blue points as field galaxies. }
		    \label{fig:caustic}
	    \end{center}
	\end{figure}

    The redshifts from the three surveys are combined to create a final redshift catalogue of the A\,1758 field, and a total of 536 cluster members were identified as lying within the trumpet-shaped caustics in the redshift vs. cluster-centric distance diagram (Fig. \ref{fig:caustic}). For this application we use the cluster centers (A\,1758\,N/S) as defined in \cite{Haines2018-re} based on the XMM-Newton image.
    While the Hectospec instrument very efficiently samples the infall regions of A\,1758, it has difficulty in placing the fibers close enough together to sample well the high-density regions of the cluster cores, so the dense sampling within these GMOS fields is particularly useful to equalize the redshift completeness across the full region.
	
	\section{Results}\label{ch:results}
	In this section we give an overview on the properties of the hot ICM, showing temperature and metallicity structure, and then reporting the strength of the cold front in the northern cluster, as well as the search for shock fronts in both, A\,1758\,N and A\,1758\,S. We show that the archival GMRT radio data confirms the previously reported radio emission in the northern cluster, but also shows strong indications for a radio halo at the position of the southern merging cluster. Additionally, we find extended emission on smaller scales in the periphery of the northern cluster.
	Finally, we examine the dynamical state of both clusters with a large number of member galaxy velocities.
		
	\subsection{X-ray results}
	\begin{figure*}
		\includegraphics[width=0.95\textwidth]{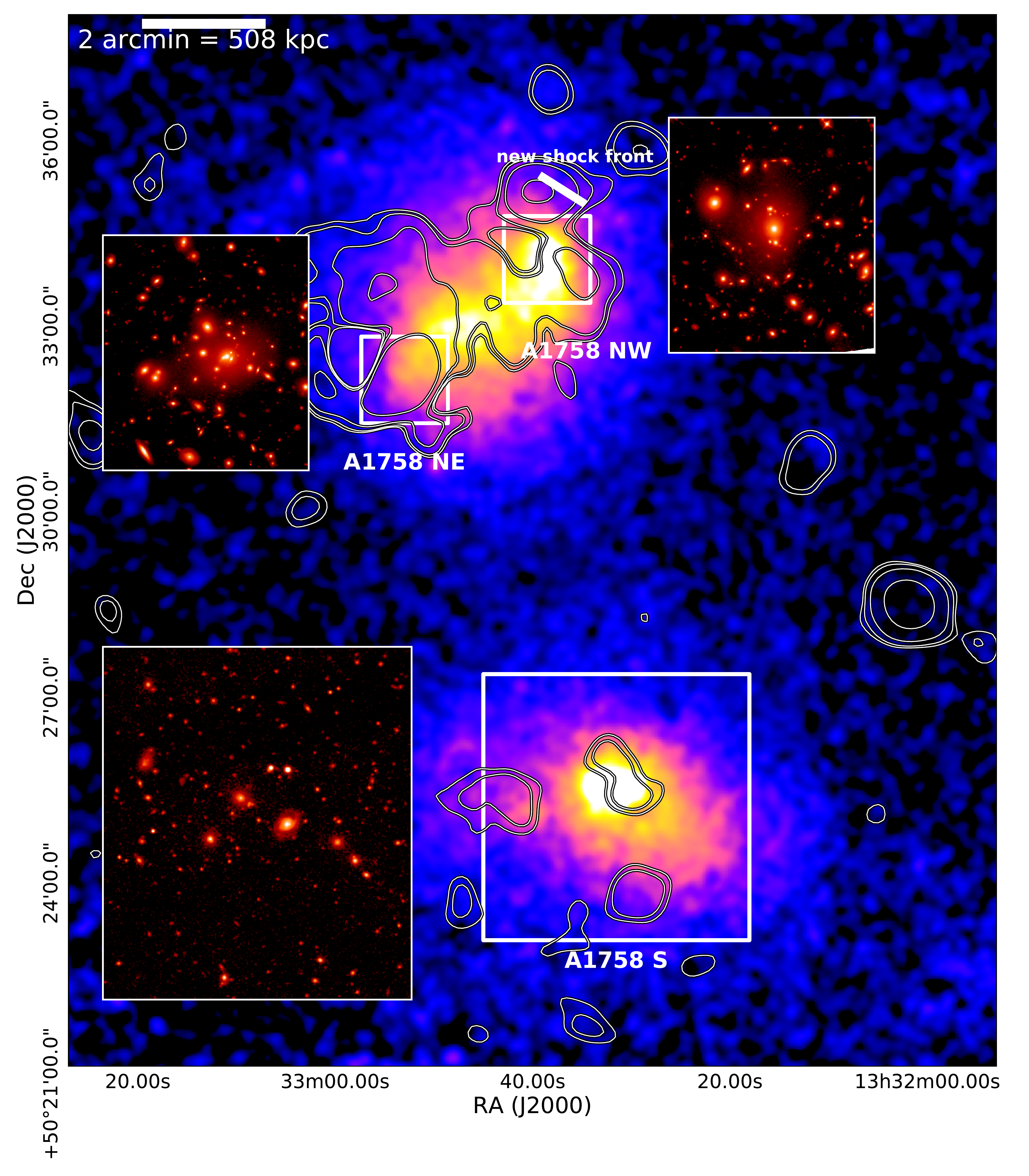}
		\caption{Exposure corrected \textit{XMM-Newton} image with GMRT $\SI{608}{MHz}$ radio contours (3,4,7 and 20 $\sigma$ levels, $\sigma=\SI{0.2}{mJy\,bm^{-1}}$), and optical closeups (top left and top right: HST observations ICZG24040 \& ICZG14040, with WFC3/F160W; bottom: SDSS).}
		\label{fig:XMM}
	\end{figure*}
	
	\subsubsection{Overall structure}
	The \textit{XMM-Newton} image (Fig. \ref{fig:XMM}) shows the two distinct clusters of Abell\,1758, the northern cluster (A\,1758\,N) consisting of two peaks ($\SI{407}{kpc}$ separation), and the southern cluster (A\,1758\,S) with a dominant peak and a smaller peak in surface brightness ($\SI{168}{kpc}$ separation). The separation of A\,1758\,S and A\,1758\,N is about $\SI{2}{Mpc}$. 
	We also find extended radio emission at the positions of the northern and southern clusters (contours in Fig. \ref{fig:XMM}), which we investigate in more detail in Section \ref{ch:radio}.
	
	\subsubsection{Hydrostatic masses}
	Although the irregular X-ray morphology and the merging interaction makes it hard to estimate a total (hydrostatic) mass, we estimate masses for the A\,1758\,S and the north-east and the north-west subclusters of A\,1758\,N. While the southern cluster is unlikely to be affected by the northern mass distribution, the NW and NE estimated masses are not independent.
	We followed the procedure of \cite{2017arXiv170505842S} to estimate hydrostatic masses (with an NFW approximation and a $c-M$ relation from \citealp{2013ApJ...766...32B} as a prior), while we excluded for each of the 3 objects (S, NW, NE) the regions of the other subclusters.
	
	We find $M_{200} = 1.06^{+0.07}_{-0.06}, 1.04^{+0.49}_{-0.25}, 1.24^{+0.10}_{-0.10}\times10^{15}\,\si{M_\odot}\mathrm{, }~$  for the S, NW and NE subcluster regions, respectively, i.e. the northern cluster is undergoing a roughly equal mass merger, while the southern cluster contains approximately the same mass as one of the A\,1758\,N subclusters. 
	The weak lensing analysis by \cite{2017MNRAS.466.2614M} was inconclusive about which of the components of the northern cluster is more massive. The authors concluded that the distortion favors the NW subcluster and magnification favors the NE subcluster to be more massive. A combined analysis gives a 79\% probability for the NW subcluster being slightly more massive. 
	We note that the X-ray temperature of A\,1758\,NE will be biased high if there are shock fronts in the ambient medium. The southern cluster has also been shown to have a comparable mass of one of the northern subclusters in past studies (Table. \ref{tab:masses}).
	Summing the two northern regions, we find a total mass of $2.3^{+0.3}_{-0.6}\times10^{15}\,\si{M_\odot}$. The total system of A\,1758 will eventually merge into a cluster with a mass $3.3^{+0.4}_{-0.6}\times10^{15}\,\si{M_\odot}$. Since the cluster is expected to grow also from accretion, this mass can be considered as a lower limit. This mass is comparable to the most massive clusters in the universe (RXJ1347.5-1145, ACT-CL J0102-4915 ``El Gordo'') have masses in the range of $(2-3)\times10^{15}\,\si{M_\odot}$ (\citealp{Bradavc2008a,Jee2014}). These values are within the range of masses found in previous studies with a variety of methods (see Tab. \ref{tab:masses}). However, the system is clearly not in hydrostatic equilibrium, so the numbers derived from X-ray studies alone have to be reconsidered in this light.
	
	Since the merger in the northern cluster is already post-core passage, another consideration can be to treat the northern object as one single cluster, and not the sum of two subclusters. In this case we estimate the total mass of A\,1758\,N by summing up the fraction of masses in the volume that the western and eastern part occupy alone. The overlapping region (71\% of the total volume) is then counted only once (and not twice as before), which results in an $M_{200}^\mathrm{N} = 1.3^{+0.3}_{-0.1}\times10^{15}\,\si{M_\odot}$. The southern cluster is clearly separated and does not need this treatment. However, we want to  stress that both clusters, A\,1758\,N and S, are clearly not in hydrostatic equilibrium.

	\subsubsection{Temperature and abundance maps}
	\label{ch:tmaps}
	The important questions addressed in this section are: How much gas is shock heated and where is it located? Is there any interaction (e.g., with shocked gas) between A\,1758\,N and A\,1758\,S? The imprints of the shocks in the ICM can be identified as surface brightness edges in the X-ray images. 
	These issues can be better understood by studying detailed temperature, abundance maps and surface brightness gradient maps.
	
	We derived temperature and abundance maps from the \textit{Chandra} data (excluding observation ID 2213) using two approaches: The first method (``contour binning'', \citealp{2006MNRAS.371..829S}) creates regions for the spectral fitting based on a signal-to-noise threshold following the surface brightness. Since spectral changes (e.g., in temperature) are often connected with changes in surface brightness, this method is expected to produce meaningful regions inside which the thermodynamical and chemical properties do not change significantly. The second method is based on an adaptive binning technique, and has been demonstrated and described, e.g., in \cite{2011MNRAS.411.1833O}. 
	The adaptive binning technique does not rely on the assumption that the temperature and metallicity structures follow the surface brightness distribution. 
	Since the spectral extraction region (selected based on a signal-to-noise and minimum counts threshold) can be larger than the actual grid size, the final map looks smoother, but the image pixels are not statistically independent.
	The parameters for the spectra of the abundance maps are a minimum signal-to-noise ratio (SNR) of 50 (fitting both temperature and abundance), while for the temperature maps we froze the abundance to the value of the abundance maps and chose an SNR of 30.
	Spectra were created from the \textit{Chandra} observations using the \verb|specextract| task, and combined afterwards (the three different observations, using the CIAO task \verb|combine_spectra|).
	\verb|Xspec| fitting allowed us to use the \verb|pgstat| statistic since we subtract the background (from rescaled blank-sky events files). ISM absorption was taken into account by the \verb|phabs| model ($N_\mathrm{H} = \SI{1.05e20}{cm^{-2}}$, \citealp{2013MNRAS.tmp..859W}), while the cluster emission was modeled by an \verb|apec| component with the \cite{2009ARA&A..47..481A} abundance table.
	
\begin{figure*}
		\includegraphics[width=0.95\textwidth]{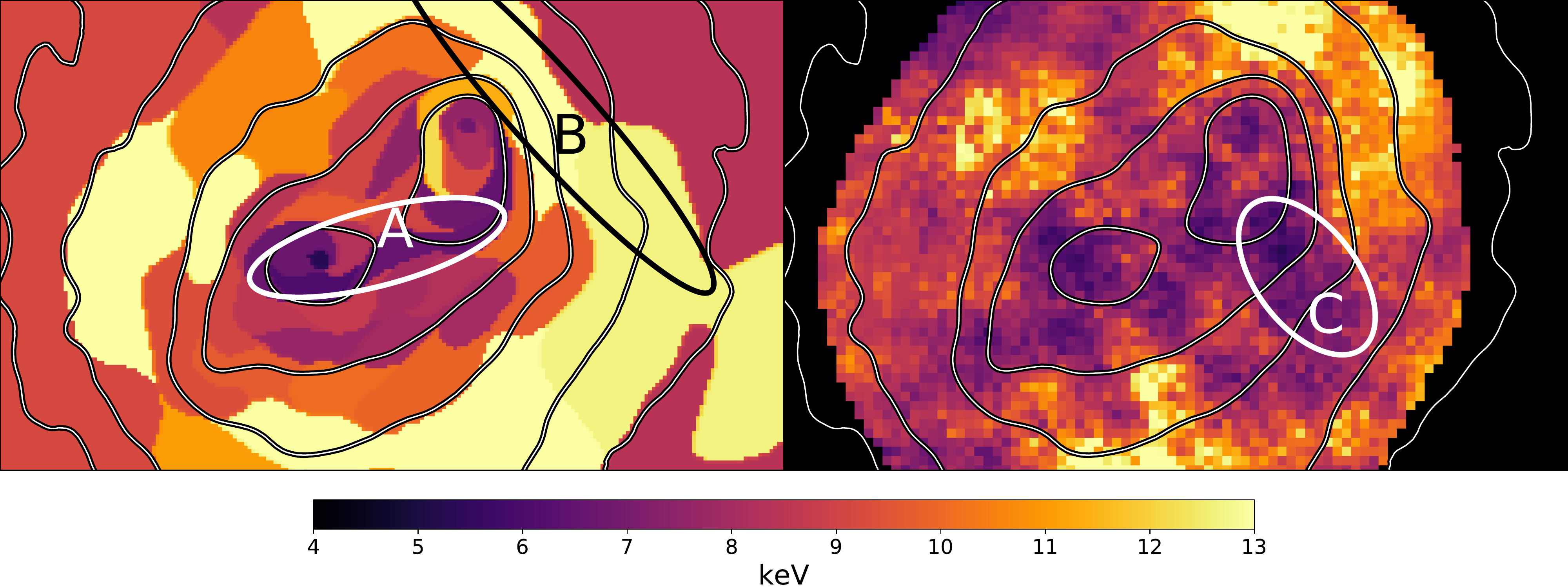}
		\includegraphics[width=0.95\textwidth]{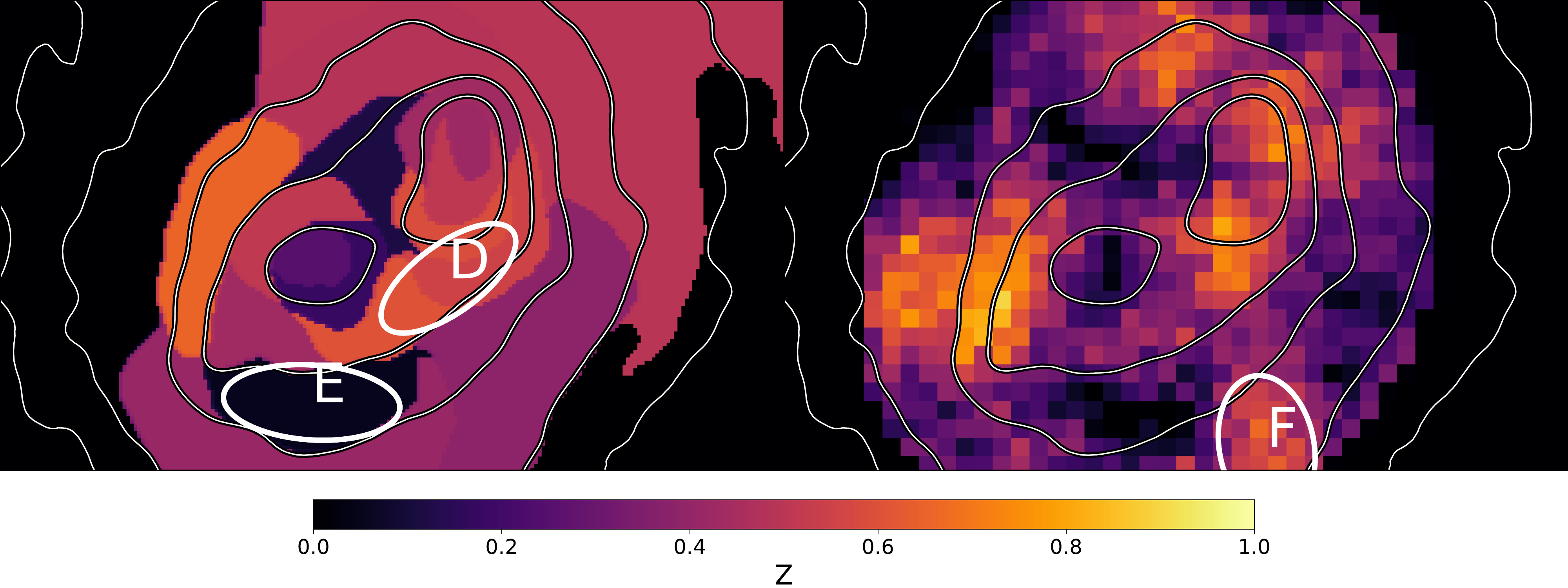}
		\caption{Abell\,1758\,N temperature (top panel) and abundance maps (bottom panel) from the \textit{Chandra} data with SNR 30 and 50, respectively. The contbin-method is used in left panels, the adaptive maps in the right. Contours show the smoothed surface brightness distribution. Labeled regions are explained in Section \ref{ch:tmaps}.}
		\label{fig:tmap_north}
	\end{figure*}

	Both temperature maps of the northern cluster are shown in the top panel of figure \ref{fig:tmap_north}, and similar features can be detected in them: Firstly, a cool, $\sim 6\,$keV, elongated region labelled (A) is located in the center, comprising both BCGs/cores. This region  extends to the south. Secondly, a hotter, shell-like region surrounds the two cool cores in the northern cluster, with temperatures up to $13\,$keV. The region (B) is part of this hot shell. 
	A temperature discontinuity is located north-west of the A\,1758\,NW core. Moreover, we find also a discontinuity east of the core of A\,1758\,NE. 
	Only the adaptive map shows a trail of cold gas to the SW (region C ). In order to eliminate the iron bias (e.g., \citealp{2000MNRAS.311..176B}) in the cores, we fitted two-temperature-component models. However, we do not find any evidence of a second, cooler component in the center of either subgroup.
	
	The abundance maps for the northern cluster (Fig. \ref{fig:tmap_north} bottom panel) reveal a highly enriched region (D) located between the two northern cores and to the east of the eastern core with a metallicity of $0.75^{+0.17}_{-0.15}\,Z_\odot$.
	The region to the south (F) had been found to be metal enriched (\citealp{2011A&A...529A..38D}), which can only be confirmed in the adaptive map.
	Taking spectra from that region we find indeed a slightly increased metallicity ($0.5^{+0.28}_{-0.25}\,Z_\odot$), which is significantly higher than the immediate surroundings (e.g., for a similar sized region slightly offset we find $0.04^{+0.15}_{-0.04}\,Z_\odot$). This  region does not show up enriched in the contbin maps, since the metallicity does not follow the surface brightness.
	The region south of the eastern core (E) shows a very low metallicity of ($0.03^{+0.16}_{-0.03}\,Z_\odot$).
	
\begin{figure*}
		\includegraphics[width=0.95\textwidth]{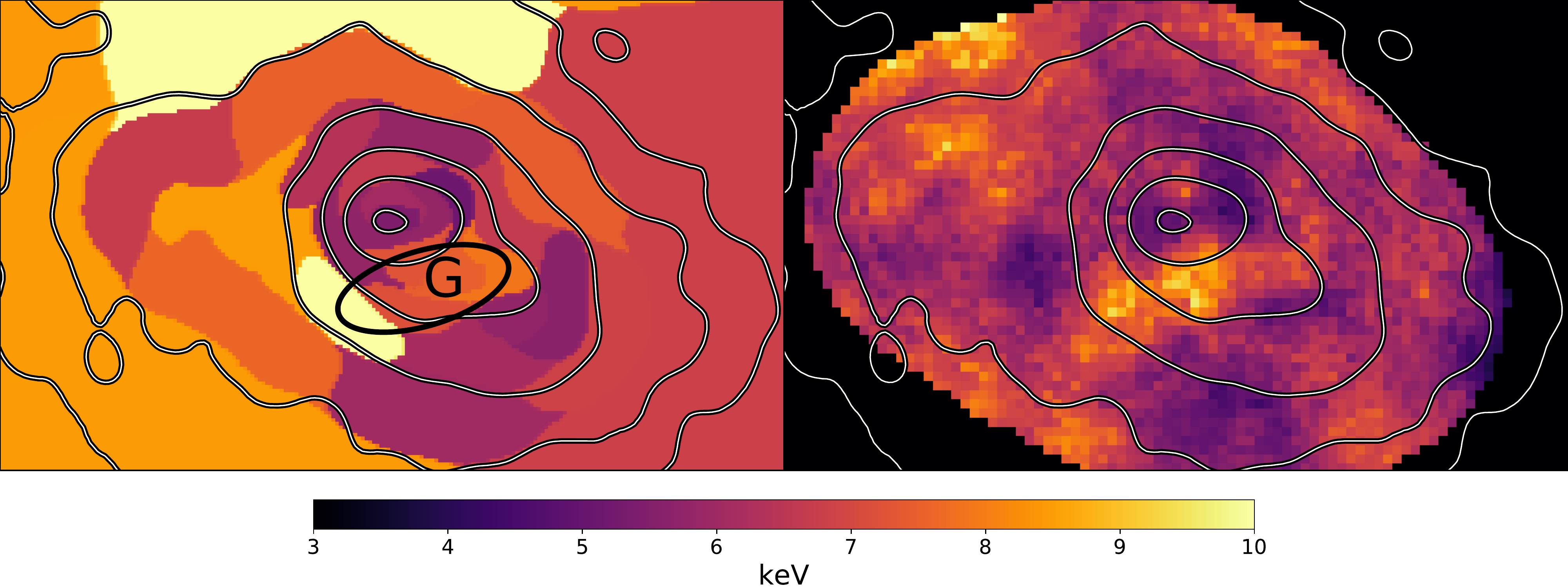}
		\includegraphics[width=0.95\textwidth]{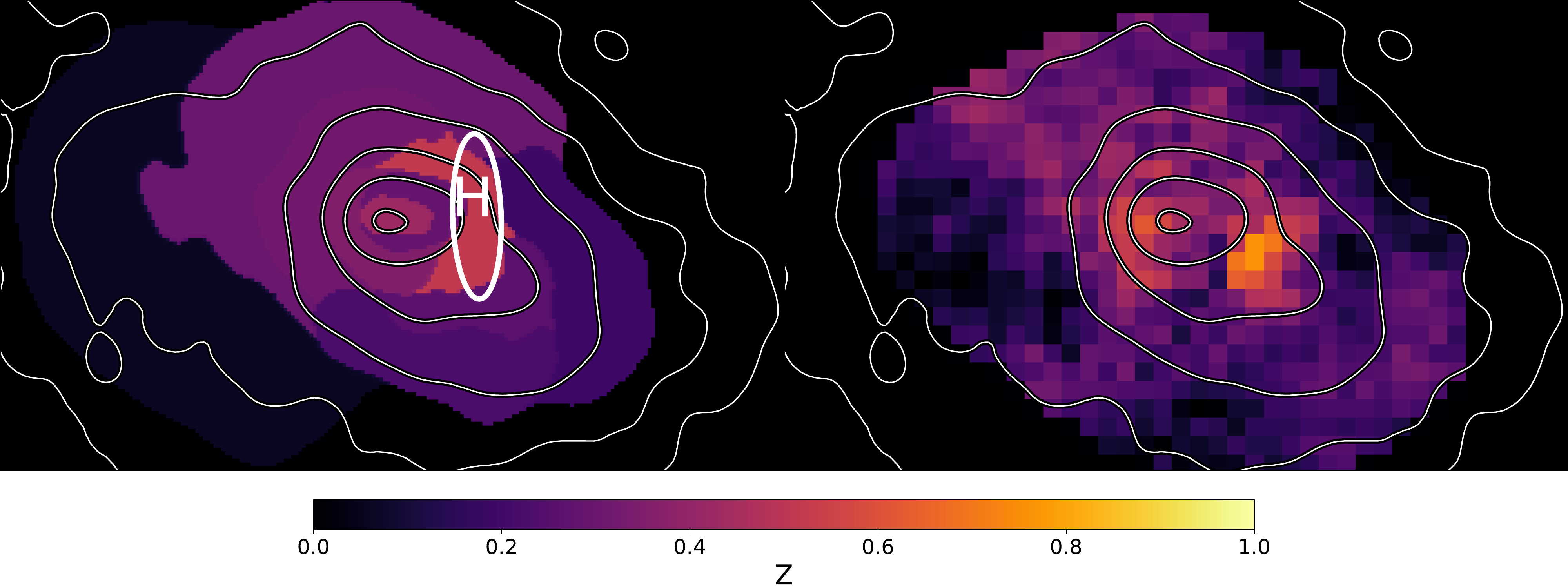}
		\caption{Abell\,1758\,S temperature and abundance maps from the \textit{Chandra} data with SNR 30 and 50, respectively. The contbin-method is used in left panels, the adaptive maps in the right. Contours show the smoothed surface brightness distribution. Labeled regions are explained in Section \ref{ch:tmaps}.}
		\label{fig:tmap_south}
	\end{figure*}

	The temperature maps for the southern cluster (Fig. \ref{fig:tmap_south} top panel) show overall lower temperatures compared to the northern cluster ($\sim \SI{5}{keV}$ in the core, and a maximum of $9-10\,$keV south of the core). The hotter region (G) overlaps with the trail of gas south-west of the core. From the temperature map it is not clear if there is hotter gas between the northern and southern cluster.
	
	The metallicity structure of the southern cluster (Fig. \ref{fig:tmap_south} bottom panel) reveals a region west of the core (H) enhanced with metals ($0.92^{+0.34}_{-0.28}\,Z_\odot$), in excess of the core region ($0.48^{+0.16}_{-0.14}\,Z_\odot$). Overall, the western outer regions are lower in metallicity than the eastern regions of the southern cluster.
    
	\subsubsection{Surface brightness edge detection}
	\label{ch:shock}
	We computed a combined surface brightness map corrected for exposure from all four available \textit{Chandra} observations (including ObsID 2213). In order to find edges and discontinuities in the surface brightness distribution, which may indicate shocks or cold fronts, we employ the Gaussian Gradient Magnitude (GGM) filtering technique (\citealp{2016MNRAS.457...82S}), which we describe in the following. 
	
\begin{figure*}
	\includegraphics[width=0.5\textwidth]{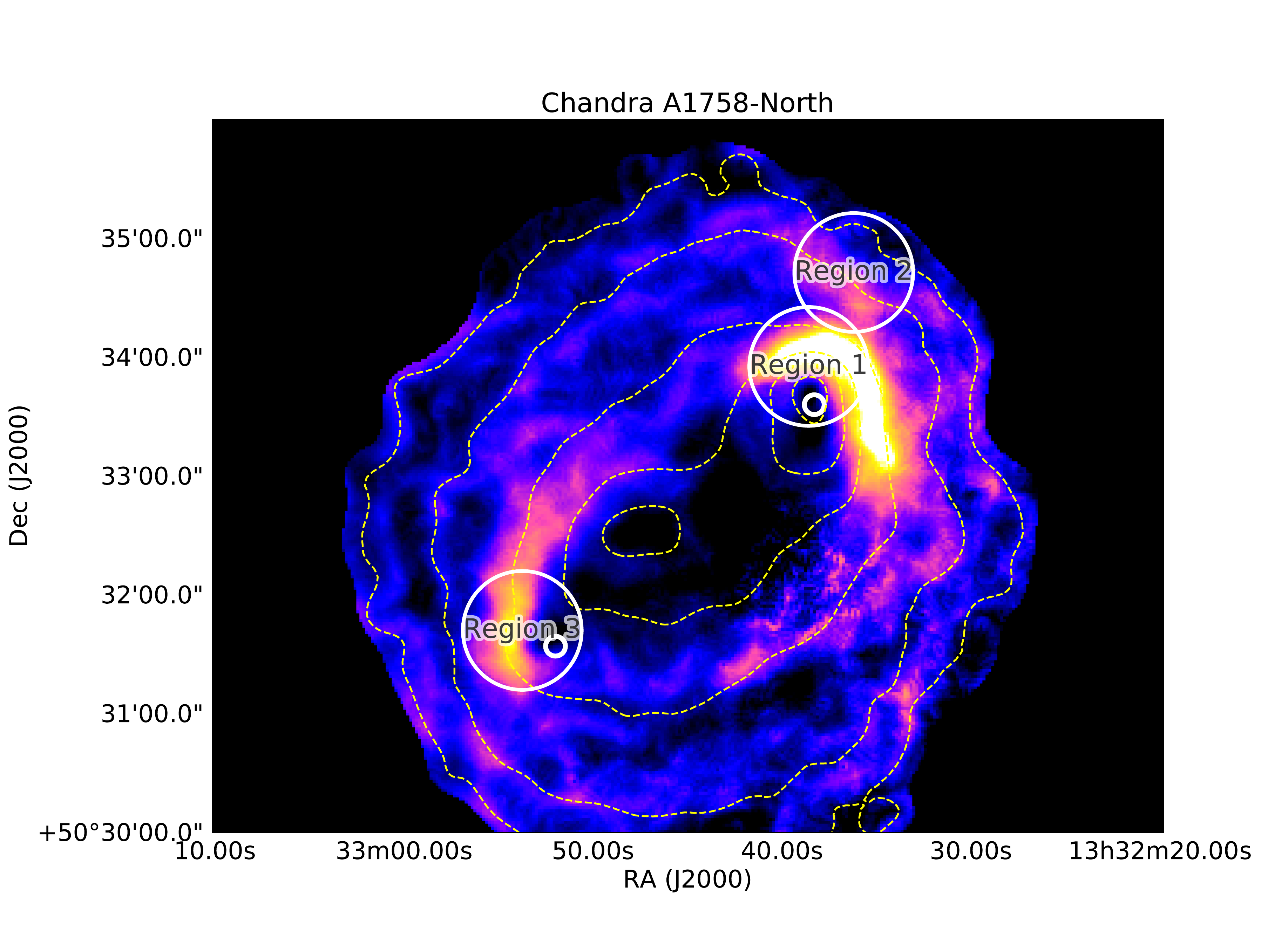}
	\includegraphics[width=0.5\textwidth]{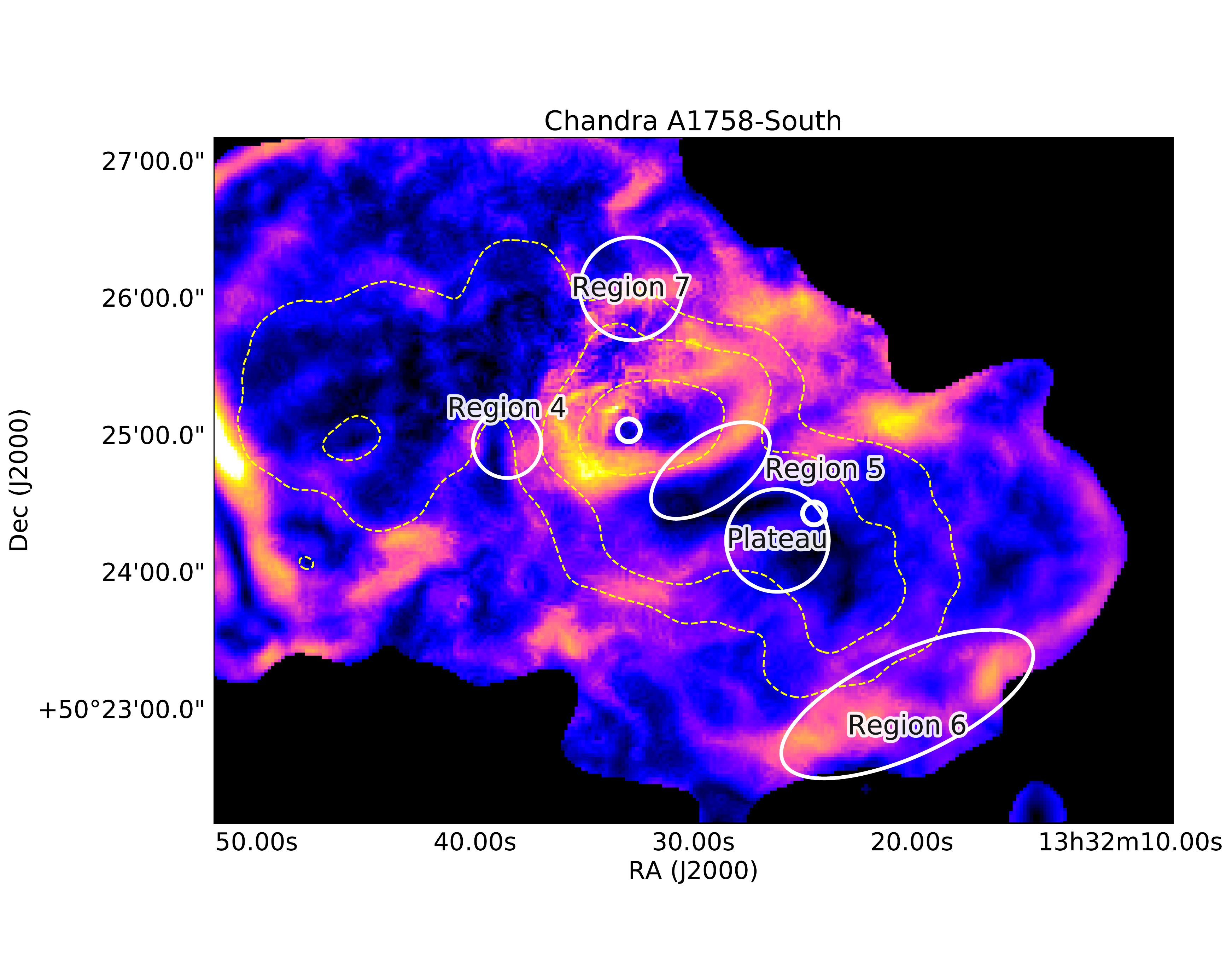}
	\caption{\textit{Chandra} Relative Gaussian Gradient Magnitude image of A\,1758\,N (left) and A\,1758\,S (right). Labeled regions are explained in Section \ref{ch:shock} and Table \ref{tab:gmm_results}.}
	\label{fig:rggm}
	\end{figure*}

	The GGM method filters spatial variations in the surface brightness on a variety of scales (through a Gaussian smoothing kernel), and combines them by weighting the scales based on the distance from a predefined center. To emphasize the discontinuities (e.g., shocks) at larger distances from the center, we normalize the GGM images by the smoothed surface brightness (see Fig. \ref{fig:rggm}).
		
    We emphasize here that the features that can be seen in the filtered images strongly depend on the individual settings for the smoothing and weighting of each scale. For GGM, we used Gaussian smoothing scales of $\SIrange{0.5}{5}{arcsec}$. 
    After testing many different scales, this range of smoothing scales turned out to most clearly display the signatures of known discontinuities (e.g., the northern cold front), and found several new features.

\begin{figure}
	\includegraphics[width=0.5\textwidth]{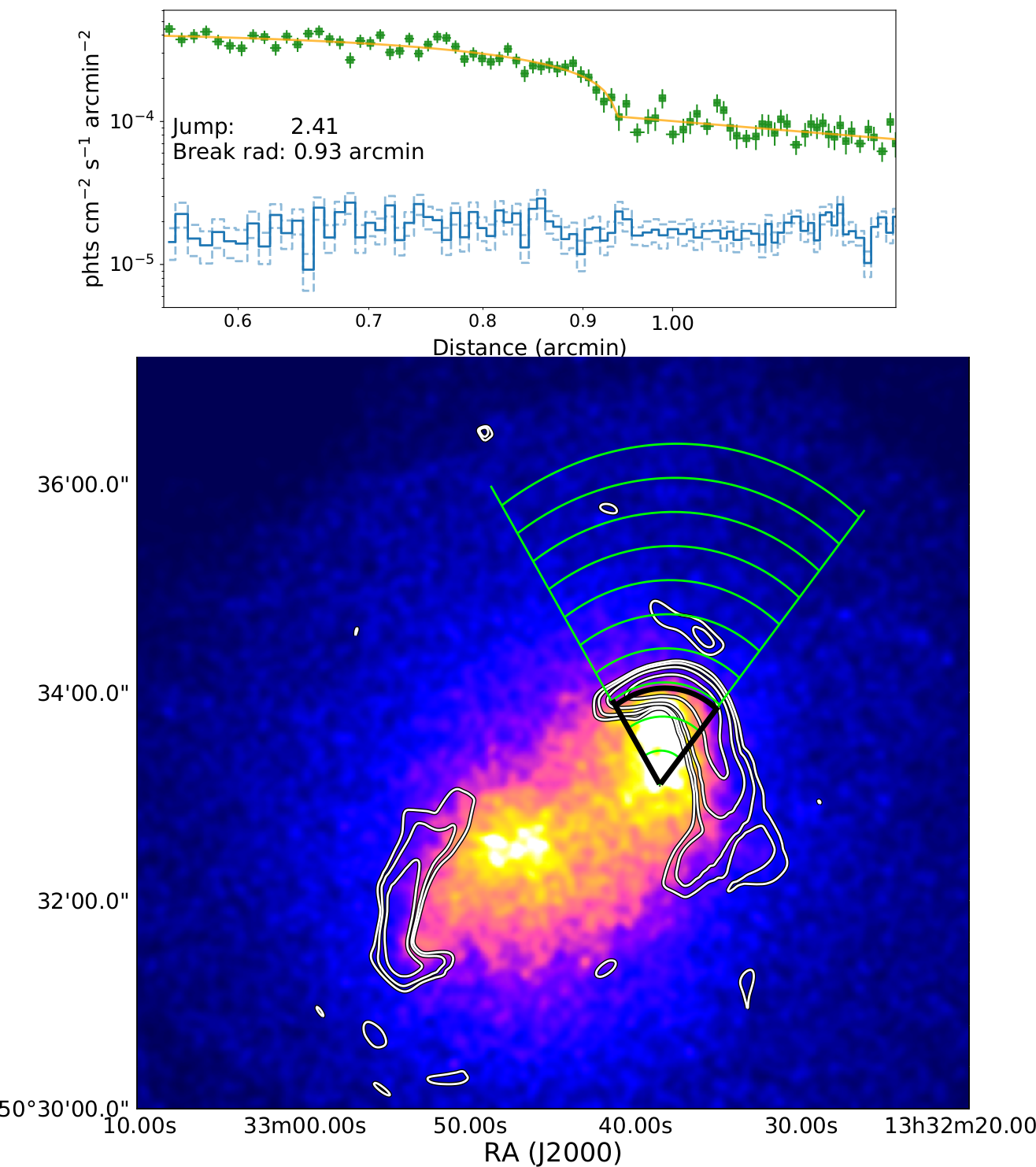}
	\caption{\textit{Region 1:} The top plot shows the one-dimensional surface brightness distribution (background in blue, background subtracted data in green, model in yellow) along the region indicated in the bottom smoothed \textit{Chandra} image in green. The contours show the edges seen in the gradient image. The thick black line shows the location of the discontinuity.}
	\label{fig:region1}	
	\end{figure}

	We list here the identified structures, to be discussed in more detail in the following:
	\begin{itemize}
		\item Region 1: The main cold front in A\,1758\,N.
		\item Region 2: Probably a shock front preceding the above mentioned cold front.
		\item Region 3: A southern cold front in A\,1758\,N.
		\item Region 4: Probably a shock front in A\,1758\,S.
		\item Region 5: A region surrounding the western half of the core with substantially lower surface brightness (it shows up bright in the GGM maps because the absolute of the gradient is shown). 
		\item Region 6: An outer edge west of A\,1758\,S.
		\item Region 7: An outer edge north of A\,1758\,S.
	\end{itemize}
	After locating shock from candidates through the gradient maps, we quantify them using the \verb|PyXel| code\footnote{\url{https://github.com/gogrean/PyXel}}, which fits a Broken Powerlaw to the one-dimensional surface brightness distribution within a given region.
	By default our region is an elliptical pie region that captures the shape of the edge.
	We show the surface brightness fits in Figs. \ref{fig:region1}, \ref{fig:region2} and \ref{fig:region3} and the resulting density jumps and Mach numbers in Tab. \ref{tab:sbredge}. In the following, we define $J$ as the temperature jump across a discontinuity ($\frac{T_\mathrm{inner}}{T_\mathrm{outer}}$), and $C$ as the density compression factor.
	\begin{deluxetable*}{cccccc}
	\tablecaption{Surface brightness edges \label{tab:sbredge}}
	\tablehead{
		\colhead{Region} & \colhead{Description} & \colhead{$C$} & \colhead{Mach$_C$} &  \colhead{$J$} & \colhead{Mach$_J$}  \\
		\colhead{1} & \colhead{2} & \colhead{3} & \colhead{4} &  \colhead{5} & \colhead{6} 
	}
	\startdata
	Region 1  & Main cold front in the northern cluster & $2.41^{+0.23}_{-0.20}$ & $2.13^{+0.28}_{-0.21}$ & $0.65^{+0.18}_{-0.08}$ & \\
	Region 2  & Outer edge north of A\,1758\,N &  $1.54^{+0.30}_{-0.25}$ &  $1.37^{+0.23}_{-0.17}$ & $1.80^{+0.70}_{-0.69}$ & $1.77^{+0.52}_{-0.66}$\\
	Region 3 & Southern edge in A\,1758\,N &  $1.53^{+0.37}_{-0.24}$ &  $1.37^{+0.28}_{-0.17}$& $1.17^{+0.18}_{-0.18}$ & $1.17^{+0.18}_{-0.19}$ \\
	Region 4 & Eastern edge in A\,1758\,S &  $3.32^{+0.48}_{-0.59}$ &  $3.82^{+3.63}_{-1.29}$ & $1.83^{+0.28}_{-0.51}$ & $1.80^{+0.39}_{-0.25}$\\
	Region 5 & Dip in the brightness &  $3.30^{+0.49}_{-0.52}$ &  $3.74^{+3.52}_{-1.13}$& $0.70^{+0.09}_{-0.08}$ & \\
	Region 6 & Far edge west of A\,1758\,S &  $2.39^{+0.49}_{-0.37}$ &  $2.11^{+0.66}_{-0.36}$ & $0.75^{+0.29}_{-0.18}$ & \\
	Region 7 & Northern edge in A\,1758\,S &  $1.46^{+0.09}_{-0.10}$ &  $1.31^{+0.07}_{-0.06}$ & $0.83^{+0.05}_{-0.04}$ & \\
	\enddata
	\tablecomments{(1) region number, (2) short region description, (3) density compression factor, (4) Mach number from density jump, (5) temperature ratio $\frac{T_\mathrm{inner}}{T_\mathrm{outer}}$, (6) Mach number from temperature jump. Mach numbers are calculated following the Rankine-Hugoniot criterion (\citealp{1959flme.book.....L})}
    \end{deluxetable*}
    
	We point out that the newly detected shock front north-west of the western core gives consistent Mach numbers determined from the surface brightness and temperature discontinuity ($\sim 1.6$).  A very weak shock can be seen in both temperature and surface brightness (with consistent Mach numbers ($\sim 1.25$) to the south of the eastern core in A\,1758\,N. On the other hand, the eastern edge in A\,1758\,S is much more visible in the surface brightness map. This could be driven by projection effects, since the merger in A\,1758\,S is not entirely in the plane of the sky.
	Regions with $J<1$ show strong surface brightness discontinuities and are likely attributed to cold fronts. 

	\subsubsection{Excess emission between the clusters}
	\label{ch:excess}
\begin{figure*}
	    \begin{center}
		\includegraphics[width=0.8\textwidth]{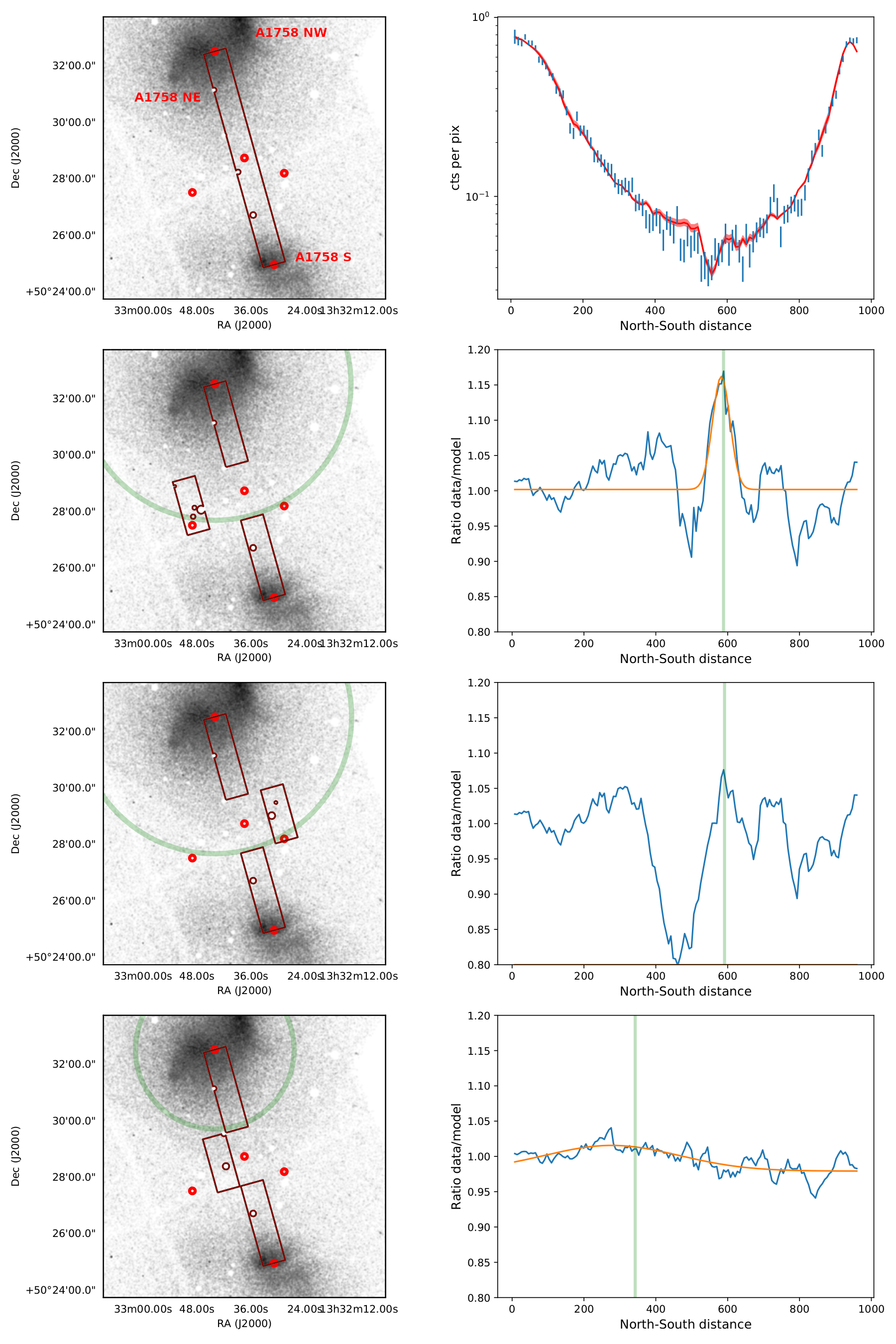}
		\caption{Modeling the cluster emission to find residuals in the inter-cluster region. 
		Left panels show the alignment of the regions (boxes), right panel show the counts per bin or ratio with respect to the model predicted counts. Red circles mark the centers of the $\beta$-models, the location of the middle between those, and the location of structure along the line of sight. The green circle marks a distance that is illustrated in the right plot. The model is determined in the top panel, where the drop in the middle is due to detector features, and also enters in the model. The other panels show a shift of the central region.}
		\label{fig:new_emission}
		\end{center}
	\end{figure*}

	If the two clusters, A\,1758\,N and A\,1758\,S, were approaching each other, one expects the gas in between to become shock heated, or at least to show signs of an adiabatically compressed filamentary structure (\citealp{2007PhR...443....1M}). 
	\cite{Botteon2018-yp} reported high ICM temperatures between A\,1758\,N and A\,1758\,S, indicating a shock by the approach of the two clusters towards each other.
	Furthermore, \cite{Ha2018-ou} show simulations that predict an equatorial shock from a major merger like the one in A\,1758\,N perpendicular to the merger axis.  We would also expect this shock to be present between the A\,1758\,N and A\,1758\,S.
	In order to find any indications of interaction between the two clusters, we carefully analyzed the surface brightness distribution in this intermediate region. 
	For this purpose we first chose a rectangular (50$^{\prime\prime}$-width stripe) region between the centers of A\,1758\,N and S, and placed on each center a $\beta$-model (see Fig. \ref{fig:new_emission} top panel). The fitting follows a Bayesian MCMC-approach to model the 2D-count distribution with a Poisson likelihood. Our model includes particle background (estimated from smoothed stowed-observations), sky background (modeled as a constant), and the exposure of each pixel. Given the best-fit model for the central region, we varied the aperture to detect residuals in the emission perpendicular to the line connecting the northern and southern clusters. We emphasize here that this approach does not model the full surface brightness distribution from all cluster components, since the system is highly disturbed and not azimuthally symmetric. But we are able to quantify in a spatial window between the clusters any excess emission, which is inconsistent with a model of the nearby cluster emission.
	
	We ran 4 MCMC chains in parallel with each having $\num{500000}$ steps, in order to constrain the model. The chains converge well within the first $\num{100000}$ steps, which we burn. We find for the northern cluster $\beta = \num{0.60(3)}$, $r_c = \SI{275(21)}{kpc}$, and for the southern one $\beta = \num{0.51(2)}$, $r_c = 107^{+10}_{-9}\si{kpc}$. 
	Both clusters have bright substructures due to the mergers, but the region between the two is much less affected by this, since none of the mergers is in the N-S direction. We find that our $\beta$-models are a good description for the emission in this region, we quantify this in a reduced $\chi^2 = 1.6$ with $\mathrm{DOF} = 91$. 
	After freezing the model we change the aperture in order to cover different regions between the two clusters.	
	We first focus on a region $\SI{50}{arcsec}$ offset to the east of the original stripe used to constrain the model. We don't detect any significant differences with respect to the model (Fig. \ref{fig:new_emission} top). Instead, if we move our mask to a region where we see several galaxies assembling an infalling subgroup of A\,1758\,N, we do detect a clear peak in the surface brightness with respect to the model (Fig. \ref{fig:new_emission} 2nd from top). If we move the mask to the west, we also detect in one region a fluctuation in emission (Fig. \ref{fig:new_emission} 3rd from top), but for most other regions we do not see any local differences relative to the reference model, while we do see excess/lack of emission on larger scales, where the merger substructure influences the model accuracy.
	
	In summary, the X-ray analysis shows that the hydrostatic derived masses fitted by an NFW profile agree well with most values from literature, even some weak-lensing studies. This is a bit surprising given that both systems are highly disturbed and clearly not in hydrostatic equilibrium. In the temperature map, the cool cores of the northern cluster subcomponents can be identified, with a stream of colder (but not as cold as the cores), enriched gas between them. In the south, the cores are not as clearly seen, but instead there is a larger diffuse region of cooler gas. The northern cluster is located within a hot shell of shocked gas that almost surrounds the entire cluster. The surface brightness analysis shows the location of several edges, two of which are shock fronts. Taking into account substructure between the two clusters, we do not find any evidence for excess emission.
	
	\subsection{Radio results}
	\label{ch:radio}
\begin{figure*}
	\includegraphics[width=\textwidth]{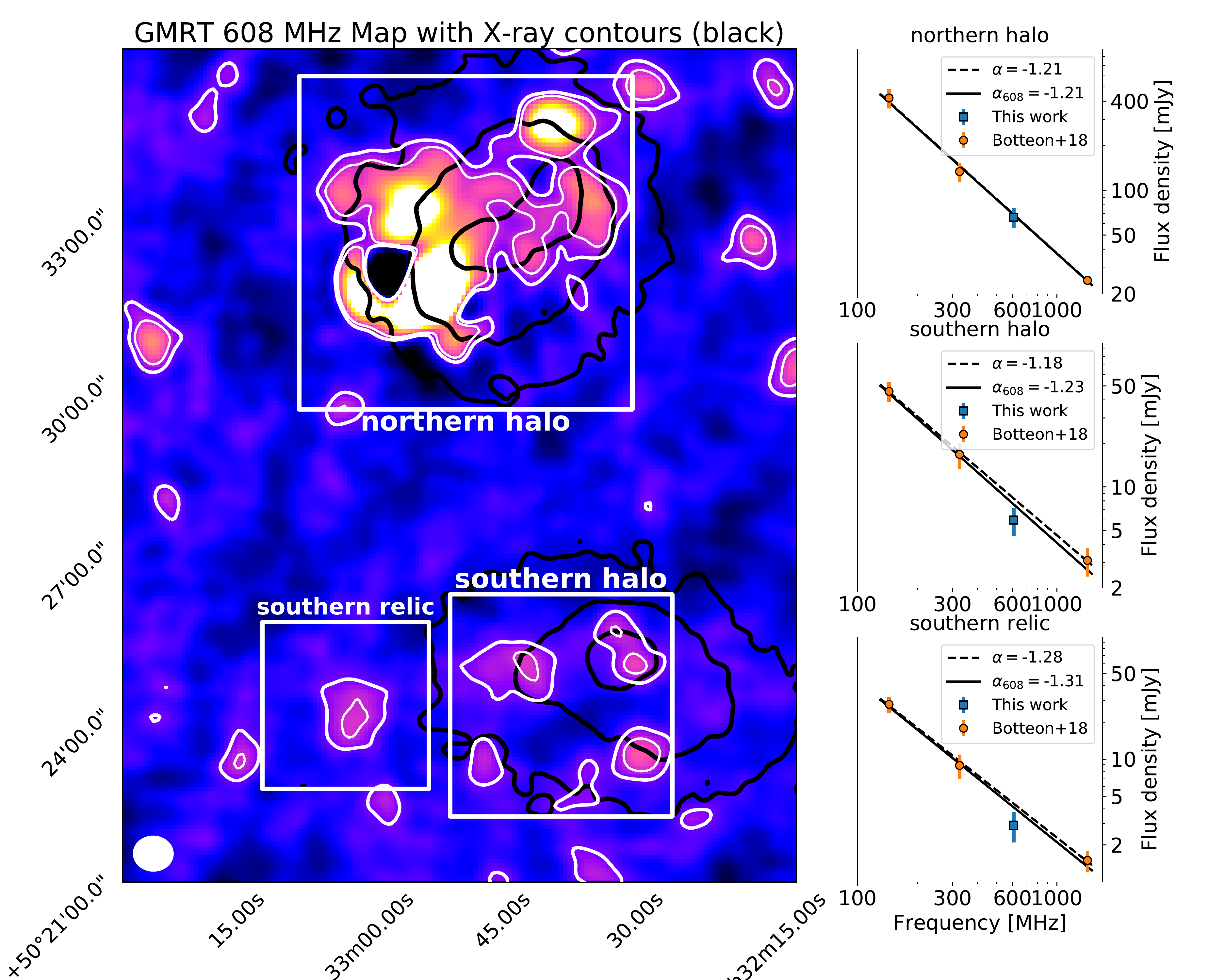}
	\caption{GMRT $\SI{608}{MHz}$ radio map (left) and radio spectra (right) for the northern halo, the southern halo, and the southern relic. The map shows 3 and 5$\,\sigma$ radio contours in white ($\sigma = 0.2\,\mathrm{mJy}$) and X-ray contours from \textit{XMM-Newton} in black. Each spectrum shows the radio spectral fit using the measurements from \cite{Botteon2018-yp} based on LOFAR, GMRT and VLA data (dashed line) and the fit including the new measurement at $\SI{608}{MHz}$ (solid line).}
	\label{fig:radio_results}
	\end{figure*}

	From the 10-hour $\SI{608}{MHz}$ GMRT observation we are able to confirm the radio halo in A\,1758\,N. Diffuse radio emission in the northern part of A\,1758 has been reported in the past (\citealp{2009A&A...507.1257G,2013A&A...551A..24V,Botteon2018-yp}), with an average spectral index of $-1.2$ to $-1.3$. We measure a flux in the point-source subtracted image of $\SI{66(10)}{mJy}$ within the 2$\sigma$ significance contour of the northern halo emission, which is consistent with interpolated spectra derived from the references above (see Fig. \ref{fig:radio_results} top panel). Taking into account our measurement at $\SI{608}{MHz}$, we confirm the spectral index of $-1.2$. 
    The morphology appears very similar to the published GMRT images in \cite{Botteon2018-yp}, but we see a brighter patch of emission to the north-west of the northern halo, at the location of the shock front.

    \cite{Botteon2018-yp} also found a radio halo at location of A\,1758\,S in their LOFAR $\SI{144}{MHz}$ data, and a relic east of it. At the same locations, we detect extended emission in our $\SI{608}{MHz}$ image, at $4.5\sigma$ significance for the halo and $4\sigma$ for the relic. Our measured fluxes are slightly below the expected powerlaw trend (see Fig. \ref{fig:radio_results} middle and bottom panels). Our regions for the flux measurements are smaller than the LOFAR detection in \cite{Botteon2018-yp}, which likely explains this missing flux.
    The morphology of the southern halo seems to consist of two peaks (east/west), while the eastern peak is centered on the X-ray emission elongated from the southern core to the east. 
	We do not see any X-ray emission at the position of the southern relic.
	
	We found extended radio emission around the infalling subgroup about $\SI{1.5}{Mpc}$ west of A\,1758\,N (Fig. \ref{fig:galaxy_density} top left, labelled G1). The group was identified by \cite{Haines2018-re}, who found it to have a mass of $\SI{4e13}{M_\odot}$. It consists of four confirmed member galaxies, all within radius of $\SI{50}{kpc}$ (projected). The radio flux is $\SI{1.5(5)}{mJy}$, and the emission is slightly extended to the south east. If this radio structure originated as the swept back radio lobes of one of the four member galaxies, we might expect it to point to the west, owing to the infall of the group toward the cluster. 

	\subsection{Structure of the galaxy population}
	\label{ch:galaxies}
	
	\subsubsection{Photometric analysis}
    For an overview of the galaxies and their spatial distribution we derived a map of the local galaxy density from the SDSS survey data (\citealp{2011yCat.2306....0A}). 
	\begin{figure}
		\includegraphics[trim={14pt 0pt 0pt 0pt},clip,width=0.49\textwidth]{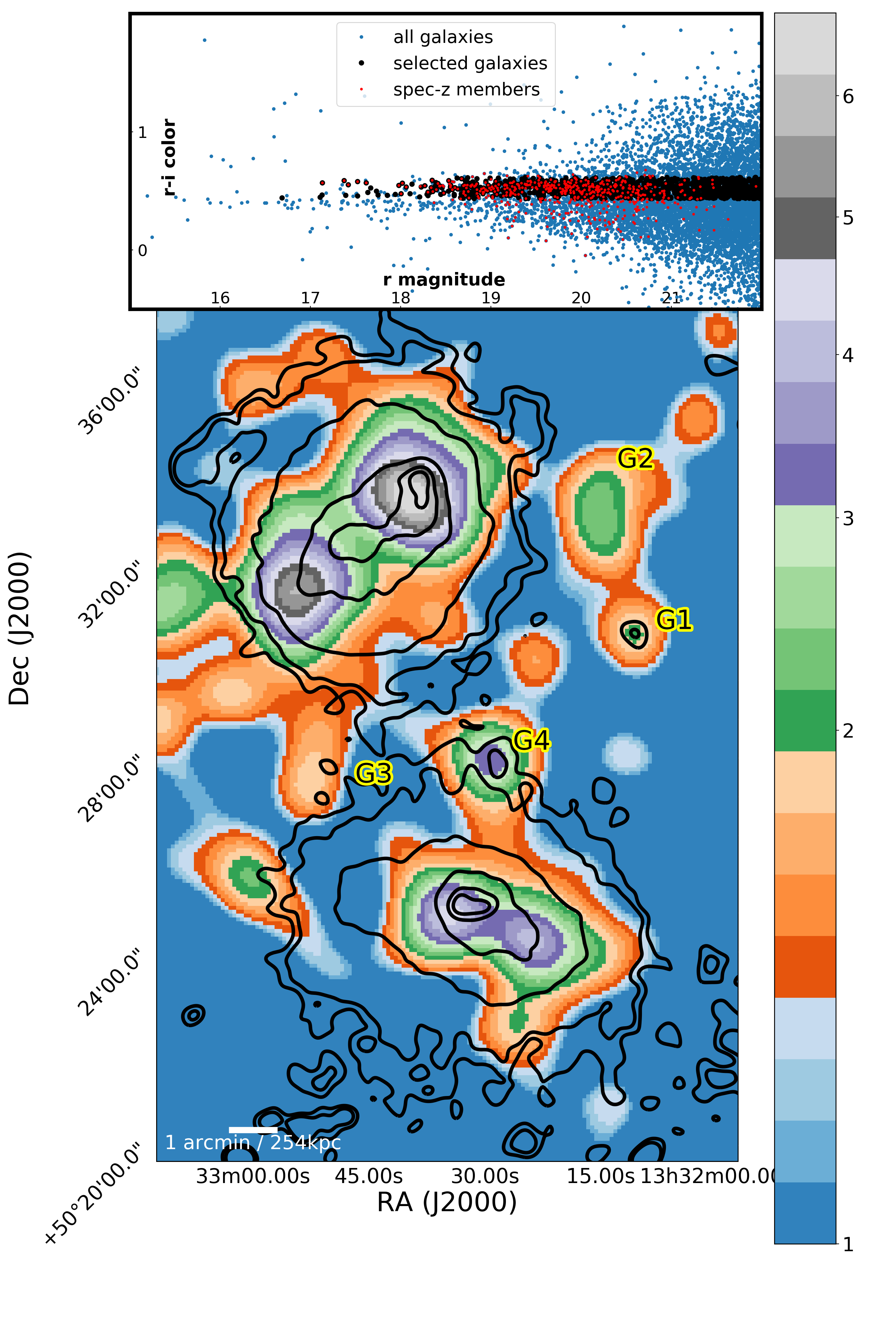}
		\caption{Top panel: SDSS color magnitude plot to identify the member galaxies (black points), using spectroscopically confirmed cluster members (red points). Bottom panel: Relative number density of photometrically identified member galaxies . The color scale is qualitative, running from the lowest density regions (blue) to the densest (green). Black solid lines are X-ray contours (\textit{Chandra} in the cores and \textit{XMM-Newton} in the outskirts of the clusters).}
		\label{fig:galaxy_density}
	\end{figure}
	We used the red sequence (Fig. \ref{fig:galaxy_density} top) to select potential member galaxies. We calibrated the selection using confirmed spectroscopic member galaxies (see Section \ref{ch:spectro_z}) and identified 1216 galaxies using the r-i band color.
	We created a density map with a Gaussian smoothing kernel ($\SI{30}{arcsec}$) that still resolves the two subclusters of A\,1758\,N, shown in Fig. \ref{fig:galaxy_density}. Note that the color scale is arbitrary, since we are only interested in the locations of galaxy overdensities and their relative intensities. We clearly see the strongest peaks around the two cores in A\,1758\,N, where the western core in the X-ray emission overlaps with the optical galaxy peak, while the eastern X-ray core lags behind the galaxies. In the southern cluster we again find two peaks in the density map (though not as pronounced as in the north), while the X-ray core is in between those peaks. In this map we also identify the infalling subgroup detected in \cite{Haines2018-re} (G1 in Fig. \ref{fig:galaxy_density}), but find more galaxy aggregations around the two clusters (G2, G3, and G4 in Fig. \ref{fig:galaxy_density} and \ref{fig:subgroup}).
	
	\subsubsection{Spectroscopic analysis}\label{ch:spectro_z}
\begin{figure}
    	\includegraphics[width=0.5\textwidth]{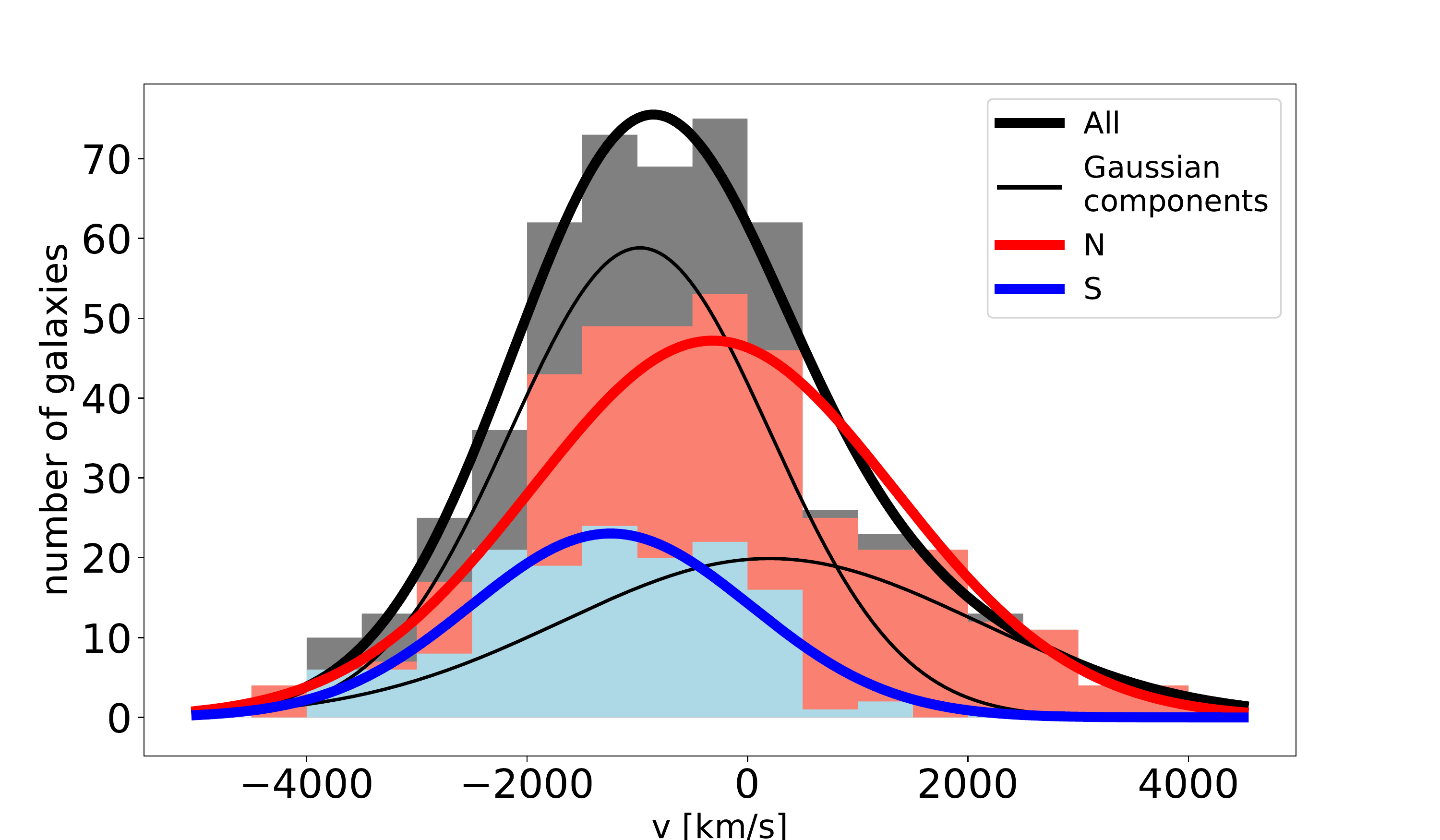}
    	\caption{Histograms of the spectroscopically identified member galaxies (grey for all member galaxies, light red for the northern, and light blue for the southern cluster). The lines represent maximum-likelihood-fitted Normal distributions to all members (black, thin lines correspond to the two components), the northern members (solid red, one component), southern members (solid blue, one component). For definitions of the regions see text. Note that $v = c \left(z - 0.279 \right)$}
    	\label{fig:redshift_histogram}
    	\end{figure}

	For a more detailed discussion we included spectroscopic information on the galaxies in our analysis. A histogram of the velocities of the galaxy population is shown in Fig. \ref{fig:redshift_histogram}. 
	We have velocities for 536 confirmed member galaxies (using the caustic method). 
	The combined, average redshift for northern and southern cluster is $0.277$. The velocity distribution is skewed (see grey bars in Fig. \ref{fig:redshift_histogram}) and is not fitted well by a single Gaussian (Shapiro-Wilk test gives a p-value of $\num{2e-6}$).
	Instead we fit two Gaussians using a maximum likelihood technique to the unbinned data (thin black lines in top panel of Fig. \ref{fig:redshift_histogram}, and the thick black line as the sum of the two). 
	We subdivide the population of member galaxies based on their position into the northern (N) and southern (S) cluster (using $\delta =$ 50d\,28m\,00s as the boundary). We obtain p-values of $\num{8e-4}$ and $\num{9e-4}$ for the northern and southern subsamples, respectively, through the Shapiro-Wilk test. 
	We see that the two Gaussian components of the total galaxy distribution (thin black lines) do not align well with the two populations (N/S, red and blue histograms and lines as single Gaussian fits in Fig. \ref{fig:redshift_histogram}).
	The northern cluster is clearly at higher redshift than the southern one ($0.278$ vs. $0.275$, about $\SI{926}{km\,s^{-1}}$ difference in velocity), and the dispersion is 30\% larger for the northern cluster. 

	\subsubsection{Gaussian Mixture Model analysis}
    \begin{figure}
		\includegraphics[width=0.45\textwidth]{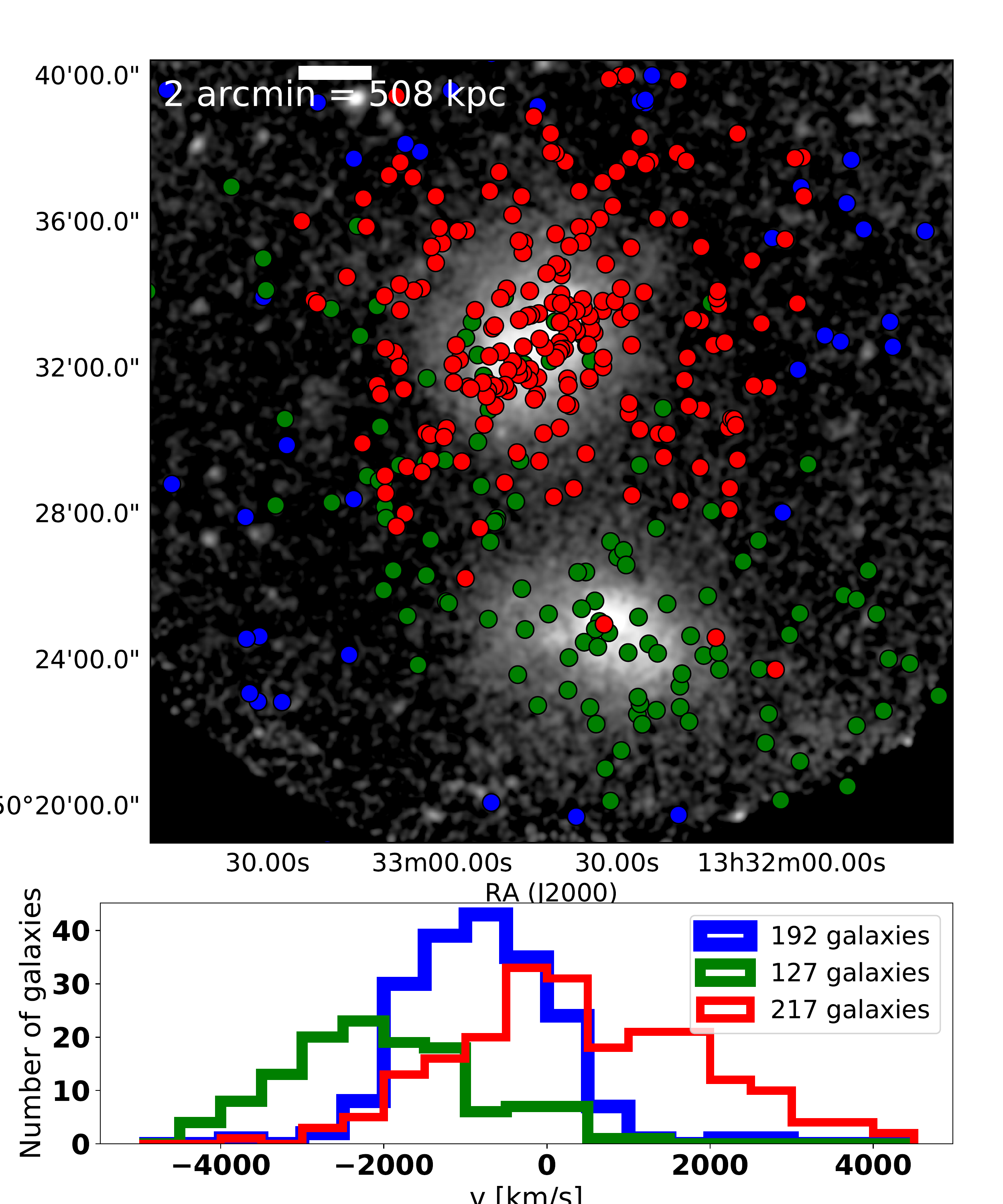}
		\caption{Gaussian Mixture Model analysis of all member galaxies of A\,1758. The top panel gives the spatial distribution of the three components found, the lower panel shows their velocity distribution.}	
		\label{fig:GMM_ALL}
	\end{figure} 	
    \begin{figure}
		\includegraphics[width=0.45\textwidth]{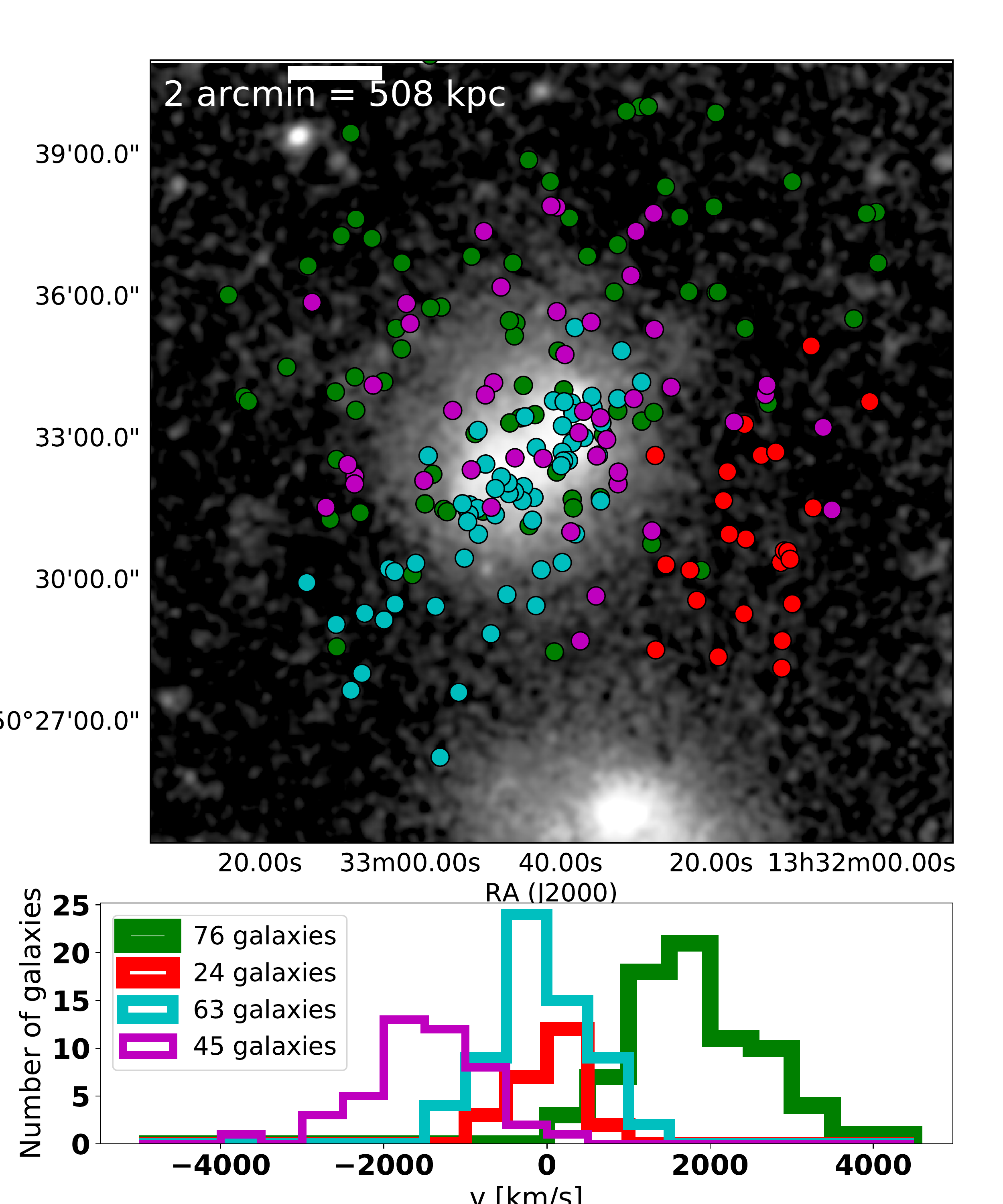}
		\caption{Same as Fig \ref{fig:GMM_ALL} except for only the northern galaxies (shown in red in Fig \ref{fig:GMM_ALL}).} 	
		\label{fig:GMM_N}
	\end{figure} 	
    \begin{figure}
		\includegraphics[width=0.45\textwidth]{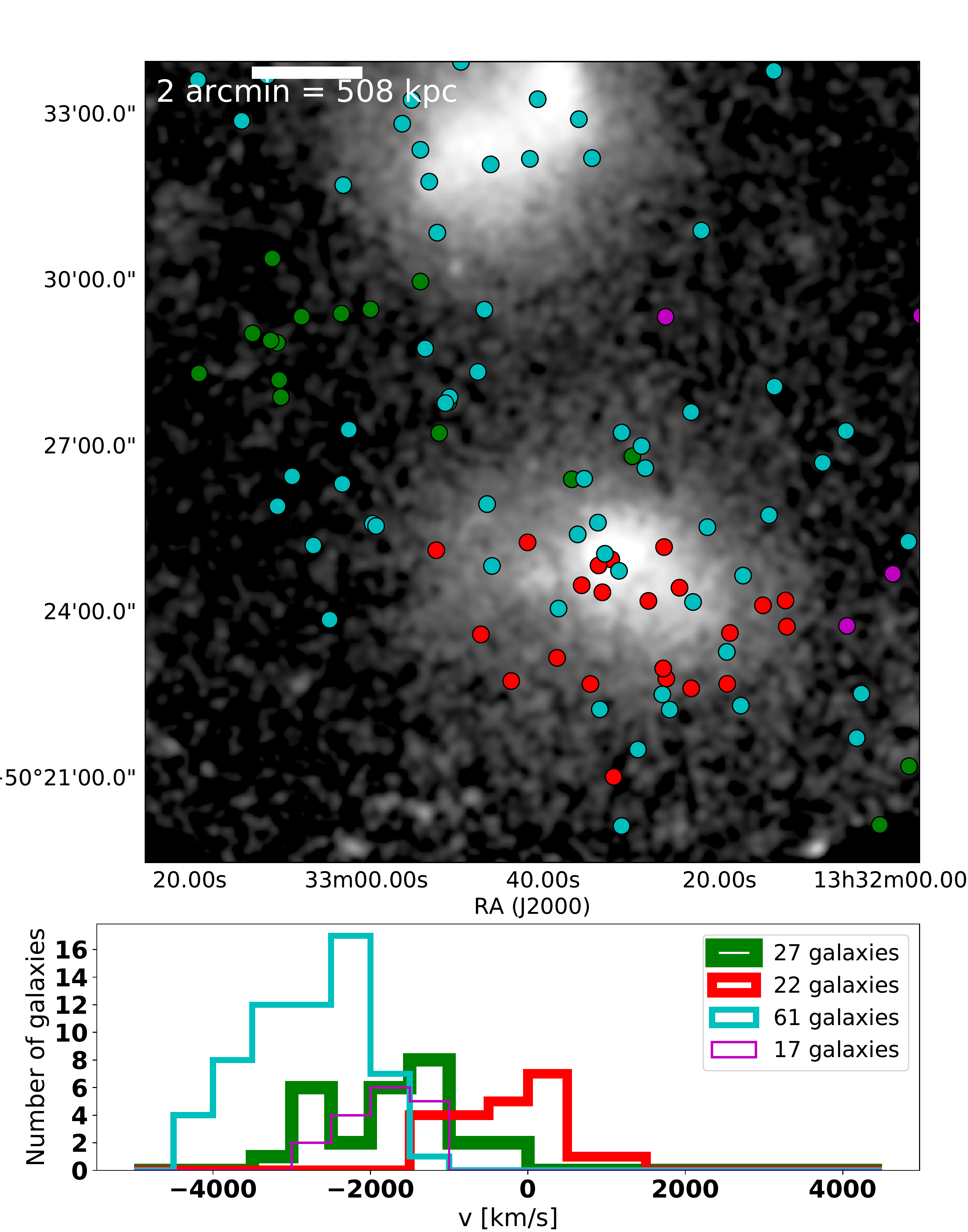}
		\caption{Same as Fig \ref{fig:GMM_ALL} except for only the southern galaxies (shown in green in Fig \ref{fig:GMM_ALL}).} 
		\label{fig:GMM_S}
	\end{figure}
	In the previous section we classified the environment of the member galaxies (North/South) based on their location with respect to the X-ray emission of the clusters. Since we are dealing with merging objects, this assumption may be over-simplistic. Substructure within the galaxy population differs from the X-ray morphology since the galaxies are effectively collisionless, and we can only observe the projected X-ray distribution. 
	\begin{deluxetable}{lcccc}
		\tablecaption{Results from the Gaussian Mixture Model\label{tab:gmm_results}}
		\tablehead{
			\colhead{Region} & \colhead{$\bar v$} & \colhead{$\sigma$} & \colhead{Fig/Color} & \colhead{$p$} \\
			\colhead{} & \colhead{km/s} & \colhead{km/s} & \colhead{} & \colhead{\%}
			}
		\startdata
		North & $\num{459(108)}$ & $\num{1593(78)}$ & \ref{fig:GMM_ALL} - red & 9\\
		South & $\num{-1998(112)}$ & $\num{1196(79)}$ & \ref{fig:GMM_ALL} - green & 31\\
		Outer ring & $\num{-797(64)}$ & $\num{863(45)}$ & \ref{fig:GMM_ALL} - blue & 1\\
		West & $\num{-11(83)}$ & $\num{381(64)}$ & \ref{fig:GMM_N} - red & 37\\
		Norther/merger &  $\num{23(73)}$ & $\num{516(54)}$ & \ref{fig:GMM_N} - cyan & 41\\
		North high-z &  $\num{1197(118)}$ & $\num{1019(82)}$ &  \ref{fig:GMM_N} - green & 1\\
	    North low-z &  $\num{-1323(109)}$ & $\num{807(81)}$ &  \ref{fig:GMM_N} - purple & 2\\
	    South high-z &  $\num{-245(146)}$ & $\num{664(114)}$ & \ref{fig:GMM_S} - red & 87\\
	    South low-z  &  $\num{-2814(108)}$ & $\num{776(82)}$ & \ref{fig:GMM_S} - cyan & 5\\
	    South separate   &  $\num{-1716(165)}$ & $\num{820(125)}$ & \ref{fig:GMM_S} - green & 17
		\enddata
		\tablecomments{Velocities are defined as $v = c \left( z - 0.279 \right)$. The last column shows the p-value of the Shapiro-Wilks test for each subsample.}
	\end{deluxetable}
	
	We use the Gaussian Mixture Model implemented in the python package scikit-learn (\citealp{scikit-learn}) to subdivide the galaxies into groups based on their 3D information, position and radial velocity.
	\cite{2017MNRAS.466.2614M} used a 1D and 2D approach to classify the galaxies in A\,1758\,N. The authors concluded that the 1D (velocity only) analysis cannot recover a bi-modality of the galaxies in the A\,1758\,NE and NW subclusters. Using the projected positions only (2D), the two subclusters can be separated, but the modelling doesn't account for the radial velocities. Here we have a larger catalog of galaxies and apply a 3D mixture model, to find an accurate galaxy classification.
	Additionally, the larger galaxy catalog not only provides more precise estimates, but also helps to separate the infalling region, and detect new groups of galaxies falling onto the system.

	In the first case we apply the algorithm to all member galaxies (see Fig. \ref{fig:GMM_ALL}, and Table \ref{tab:gmm_results} for values). We restrict the number of components to 3, since we are interested in a distinction between the northern and southern cluster, plus any additional component unrelated to the two. 
	As expected we find a distinction of galaxies within each X-ray halo (shown in Fig. \ref{fig:GMM_ALL} in red for the northern, and green for the southern halo). In the velocity histogram the two populations are more distinct than if we just split them based on a threshold in Declination (compare with Fig. \ref{fig:redshift_histogram}). The third component (blue in Fig. \ref{fig:GMM_ALL}) selects all galaxies around the two clusters, and has a more peaked velocity distribution than the the northern and southern components.
    The confined velocity spacing of this third component can be explained by the small infall velocity present at the large radii. 
    Using a larger number of initial components for the Gaussian Mixture Model does not change the split between north and south, but splits the surrounding galaxies spatially (Fig. \ref{fig:GMM_ALL} blue), which are typical for the infall regions of clusters, as galaxies at $\sim 2-5\,r_{200}$ are only moving at $\sim \SIrange{200}{1000}{km/s}$ toward the cluster, and the coherent infall compresses the distribution in redshift space (see \citealp{2007MNRAS.376.1577D,2015ApJ...806..101H}).
    
	\begin{figure*}
		\includegraphics[width=0.44\textwidth]{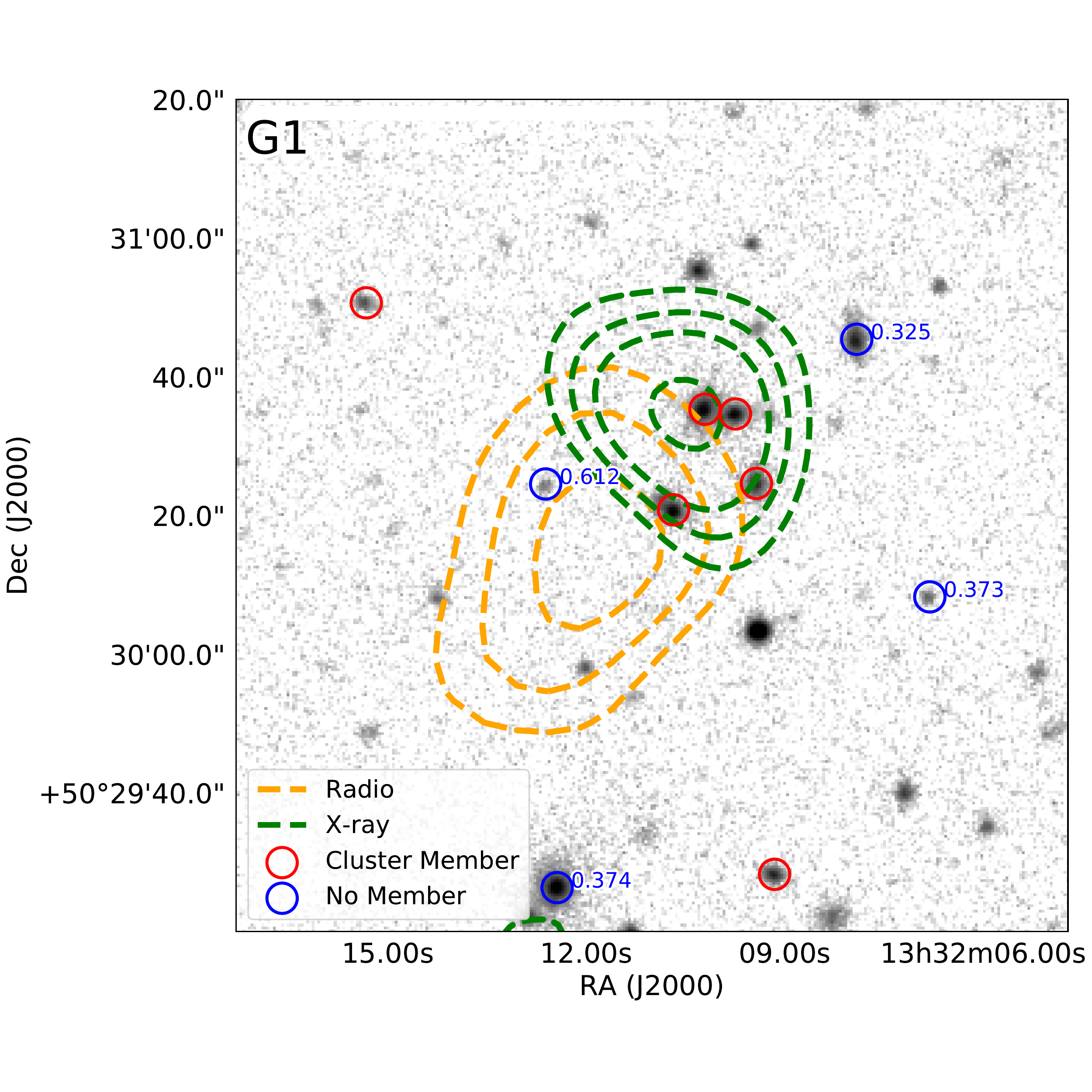}
		\includegraphics[width=0.44\textwidth]{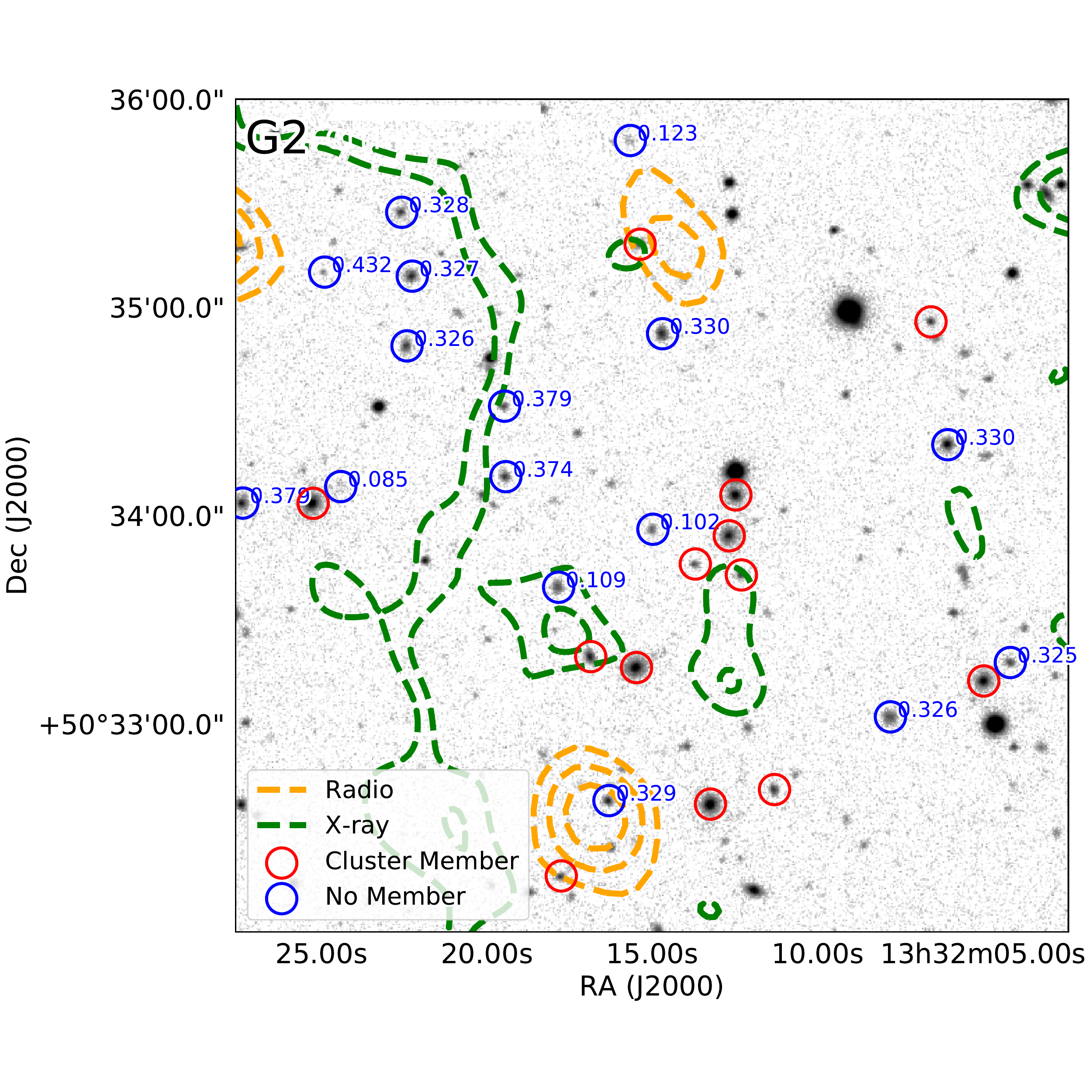}\\
		\includegraphics[width=0.44\textwidth]{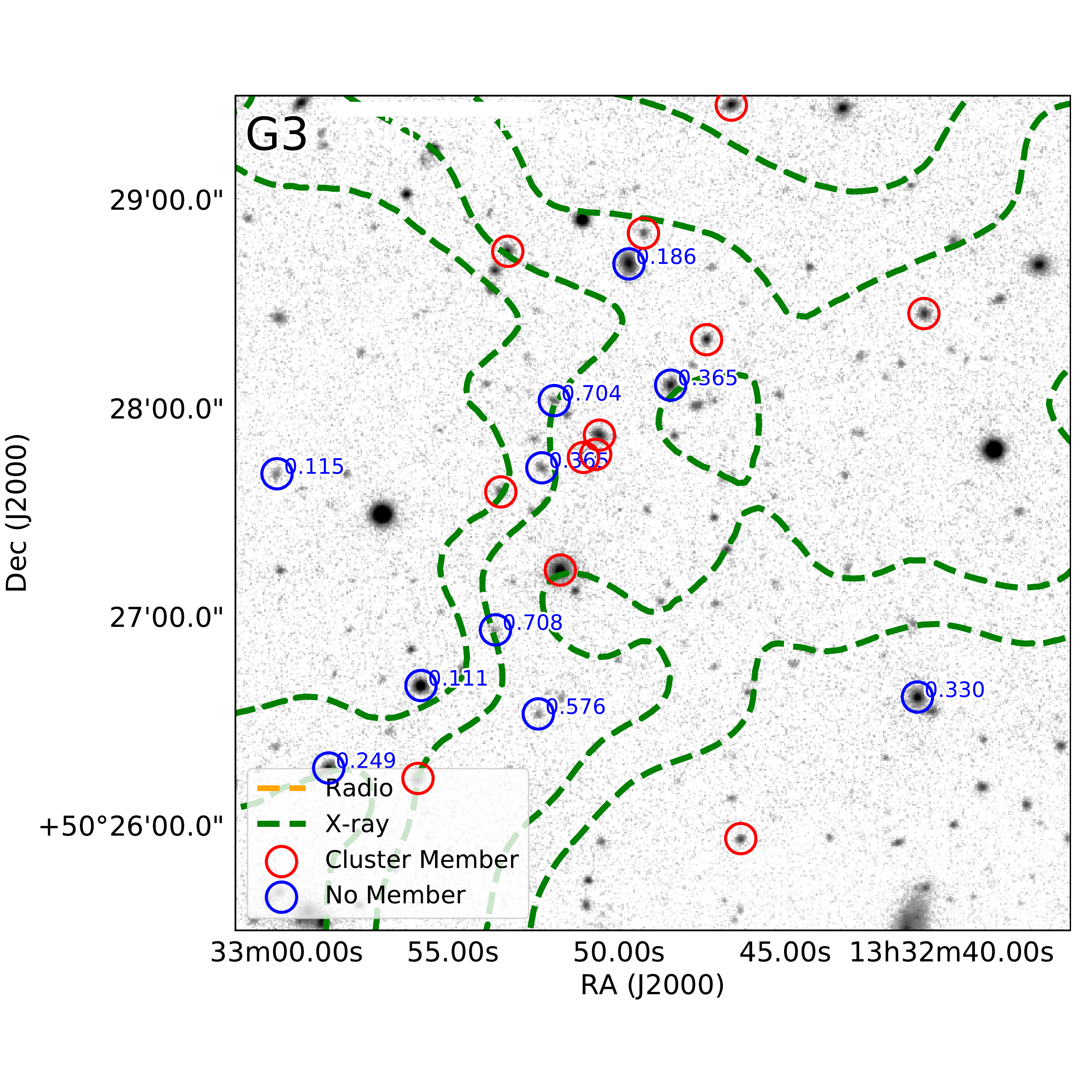}
		\includegraphics[width=0.44\textwidth]{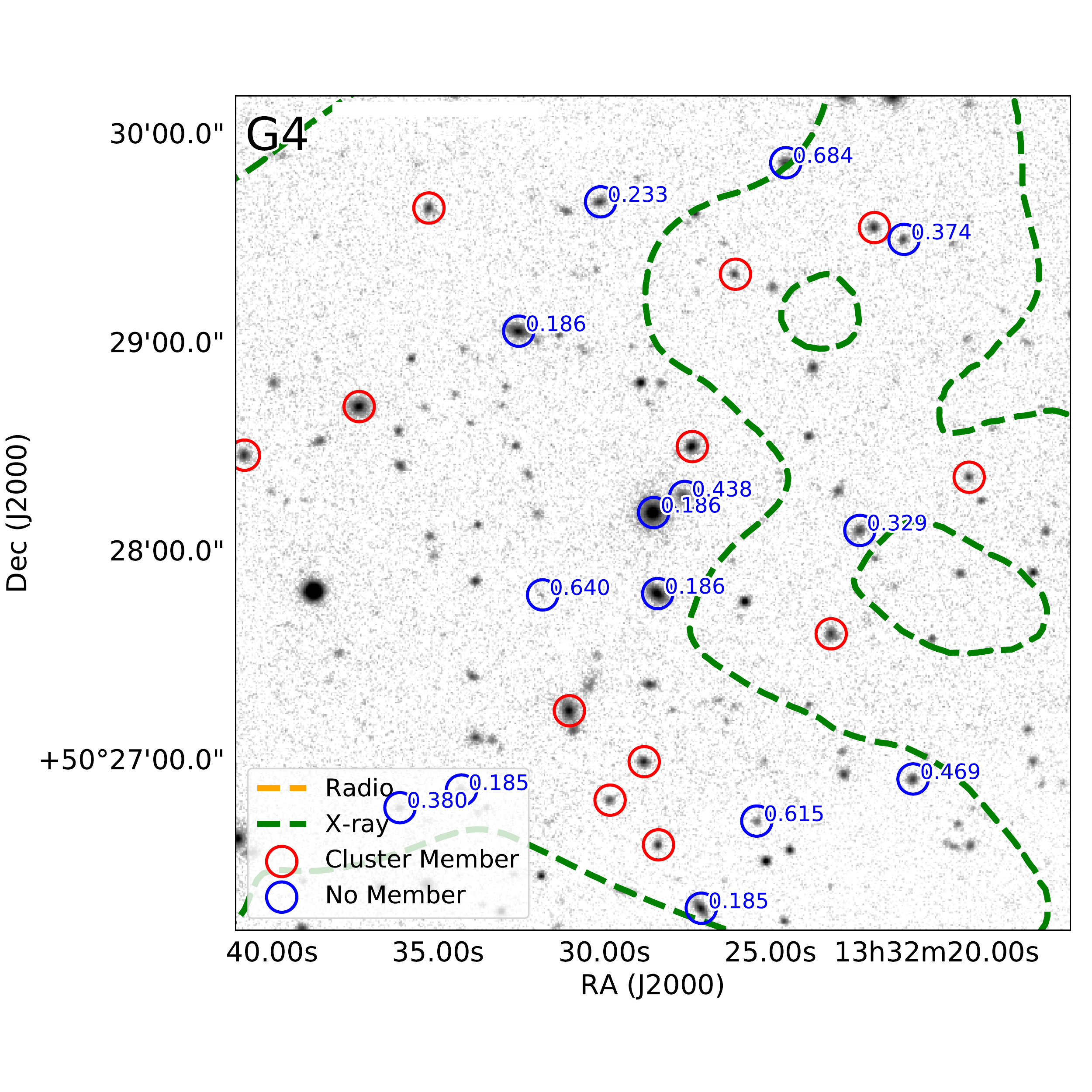}
		\caption{SDSS r-band image with Radio/GMRT (orange) and \textit{XMM-Newton} (green) contours. Spectroscopically identified galaxies are marked with circles (red for A\,1758 member galaxies, blue for others whose redshift is indicated).}	
		\label{fig:subgroup}
	\end{figure*}	
	Galaxies classified as northern or southern can now be further investigated: We applied the same algorithm to the galaxies to each of the two classes. In case of the northern galaxies (Fig. \ref{fig:GMM_N}) we find a redshifted population (shown in green), and a blueshifted population of galaxies (in purple), both located in the central and northern part of A\,1758\,N.
	We do not see a distinction between the NW and NE subclusters in velocity space, which points toward a merger in the plane of the sky. \cite{2017MNRAS.466.2614M} found the merger in A\,1758\,N to be almost in the plane of the sky (\SI{21(12)}{deg}).
	In light blue one can see a population of galaxies narrowly peaked in the velocity histogram, and spatially confined along the merger axis. 
	The green and purple galaxy populations in Fig. \ref{fig:GMM_N}, located at the extreme ends in velocity space but not spatially separated, could represent galaxies that were scattered during core passage. 
	To the west we see another assembly of galaxies (in red) with a small velocity dispersion. 
	
	In the southern cluster (Fig. \ref{fig:GMM_S}), we find redshifted and blueshifted populations  (in red and cyan, respectively)
	with very large spatial extent, and a high degree of spatial overlap. The population shown in green might either belong to an infalling group of A\,1758\,N (due to their high density), or belong to A\,1758\,N itself and be falsely classified as part of the southern cluster. The large velocity separation of the two components $(\sim \SI{2500}{km\,s^{-1}})$ shows that the merger in the southern cluster is much less in the plane of the sky than the northern merger. However, the two components in A\,1758\,S separated in velocity space have large spatial overlap, which can be explained by the merger in A\,1758\,S being in the core passage phase.
	
	\subsubsection{Associated groups of galaxies}
	Figure~\ref{fig:galaxy_density} shows the locations of four likely subgroups around the clusters (labelled G1-G4), and Figure~\ref{fig:subgroup} shows the galaxies associated with these subgroups in more detail. As described above, G1 has a relatively strong, diffuse radio component in our GMRT observation, and we see four spectroscopically confirmed member galaxies within its X-ray halo. 
	G2 is located $\SI{4}{arcmin}$ west of the western core of A\,1758\,N and comprises 8 spectroscopically confirmed members within $\SI{1}{arcmin}$. The X-ray emission of G2 is very weak. 
	G3 is located between the northern and southern cluster, and also comprises 8 confirmed member galaxies. Since the subgroup falls within the edge of the cluster emission it is hard to disentangle any X-ray component associated with G3 itself. However, the modeling shown in section \ref{ch:excess} reveals some excess X-ray emission at the location of G3. One could assume G3 is a bound system with its X-ray halo, but \cite{2008PASJ...60..345O} did not find a dark matter halo at this location. 
	The region G4 corresponds to a foreground galaxy cluster (\verb|GMBCG J203.11892+50.4966|, $z=0.195$) detected in SDSS data (\citealp{2010ApJS..191..254H}). We find 5 spectroscopically confirmed member galaxies (based on a caustic diagram) of this system within $\SI{400}{kpc}$.
	
	\section{Discussion}\label{ch:disc}
	\subsection{Radio Halos}
	Using archival GMRT $\SI{608}{MHz}$ data, we confirmed the previous finding at lower frequencies \citep{2013A&A...551A..24V,Botteon2018-yp} of a radio halo in A\,1758\,N. 
	Our estimated flux is in very good agreement with the inferred flux from the spectral index of \cite{2013A&A...551A..24V}. The morphology of the halo is patchy: Peaks of emission are found east and west of the galaxy peaks of the subclusters, along the merger axis. 
	We can also confirm the halo and relic in A\,1758\,S recently reported in \cite{Botteon2018-yp}, although with lower significance than the northern halo. We see several patches of emission in the southern halo (see box in Fig. \ref{fig:radio_results}) but the total flux is consistent with the reported values at lower frequencies, extrapolated to $\SI{608}{MHz}$.
	The resulting spectral indices of the halos of A\,1758\,N and S are almost identical, although the two clusters are in very different stages of the merger: From the X-ray and galaxy velocity information one can assume that A\,1758\,S seems to be in the core passage stage, while A\,1758\,N is far beyond that stage, see Section \ref{ch:spectro_z}. Furthermore, A\,1758\,N is much more massive, so the merger energy should be higher. It is possible that these two effects balance each other to produce a similar spectrum as in A\,1758\,S. 
	It is still unclear why the merger in A\,1758\,S has produced a radio relic, while none is detected in its northern counterpart. 
	\cite{Weeren2017} suggested that radio relics are connected with re-acceleration of relativistic particles introduced into the ICM by a past AGN outburst. In the case of A\,1758\,S we do not find any galaxy in the relic vicinity that could host a radio AGN.
	
	\subsection{ICM structure}
	Temperature maps using the the deep, high resolution \textit{Chandra} data reveal a complex thermodynamical and chemical structure of the two clusters. Overall, we find results consistent with previous findings from \textit{XMM-Newton} data (\citealp{2011A&A...529A..38D}). 
	We see an envelope of hotter gas around A\,1758\,N, especially to the north-west, in the direction of the shock front detected in Section \ref{ch:shock}.
	The two clusters, A\,1758\,N and S, are in different states of the merger, so the metal distribution provides information the redistribution of heavy elements in the ICM during a merger. 
	The metal distribution in the A\,1758\,S is more compact than in the northern cluster, and the peaks roughly follow the X-ray surface brightness. This means the cores still hold most of the heavy elements. In the northern cluster metals are more mixed: between the cores and east of the eastern core are the highest abundances, while especially the eastern core is very metal poor. 
	We also find two regions north and south of the merger axis with metallicities below 0.2, significantly lower than the surrounding medium (0.4). The origin of these metal depleted regions is not clear, and they are not observed in the southern cluster.
	
	The two northern cores hold colder gas ($\sim \SI{6}{keV}$), which is also (partly) located in the region between them. The surface brightness distribution and shape of the cold-gas around the NW core indicates a change in its direction of motion, trending to the north. The NW shock front (region 2 in Tab. \ref{tab:sbredge}) is well aligned with this inferred change of direction.

	Simulations of A\,1758\,N (\citealp{2015MNRAS.451.3309M}) predict bow shocks with high temperatures. We identify similar temperature structures in our map (using the contbin code), and also detect small discontinuities in the surface brightness distribution at larger radii, indicating that these are indeed shock fronts. In the case of the edge to the north-west (Region 2 in Tab. \ref{tab:sbredge}) we find the same Mach number from the surface brightness edge and the temperature jump ($\mathcal{M}\simeq$1.5). 
	The location of this shock front is north of the north-west core, not along the inferred merger axis (north-west). 
	This is unexpected, however the interaction between the two halos in A\,1758\,N might have changed their trajectory. 
	The mach number of the shock gives an estimate of the relative velocity, which is about $\SI{2100}{km\,s^{-1}}$. This translates into a time past core passage of $\SIrange{300}{400}{Myr}$ (using a separation of the two merger constituents in the northern cluster of $\SI{750}{kpc}$). This is in agreement with the finding of \cite{2015MNRAS.451.3309M} using hydrodynamical simulations in order to understand the system of Abell\,1758. 
	However, as shown by \cite{Springel2007-un,2015MNRAS.451.3309M} the relative velocity of the two clusters is smaller, since the shock front moves away from the cluster, and the ambient medium on the pre-shock side is gravitationally accelerated toward the shock, creating an even high relative velocity.
	
	The region between A\,1758\,N and S is of particular interest. \cite{Botteon2018-yp} reported the detection of a temperature discontinuity in one sector between the two clusters. 
	We investigated this, taking into account the presence of background/foreground clusters and infalling subgroups. We placed 5 boxes in the region of interest and match their position and size to the ones used in \cite{Botteon2018-yp}, then analyze the temperature 1) excluding only point sources, and 2) also excluding the position of the subgroups (G1-3, Fig. \ref{fig:subgroup}), and the reported foreground cluster \verb|GMBCGJ203.11892+50.46966|. We only find consistent temperatures with \cite{Botteon2018-yp} when just automatically detected point sources are excluded. The high temperature jump has been reported for the second box (from east to west), where the subgroup G3 is located. Excluding the G3 region decreased the best fit temperature by about 33\%, while the temperature uncertainty remaining large. This temperature reduction means that a shock, even a weak one, is now only $1.5\sigma$ significant, while a shock with Mach number of at least $1.5$ would be only $1.1\sigma$ significant. Using the published temperatures in  \cite{Botteon2018-yp} (i.e., not excluding the additional regions), a shock with $\mathcal{M}>1.6$ has $1.7\sigma$, while a weak shock is $2.3\sigma$ significant.
		
	\subsection{Infalling subgroups}
	\label{ch:infalling_subgroups}
	As shown in Section \ref{ch:galaxies} and Fig. \ref{fig:galaxy_density} we also find several nearby aggregations of galaxies or small groups of galaxies, possibly infalling onto the clusters A\,1758\,N/S. These objects are identified through local galaxy overdensities, since their X-ray counterpart is too faint in two of the three. In the case of the previously reported group G1 (\citealp{Haines2018-re}), we are able to detect the X-ray emission, and even see a radio tail to the south-east. Since compact sources have been removed in uv-space from the radio image, we can assume that there is diffuse radio emission associated with this galaxy group ($\SI{1.5(5)}{mJy}$). However, the detection significance is at the $3\sigma$ threshold and fluxes should be checked at other frequencies. We can see the source in the LOFAR image published in \cite{Botteon2018-yp}, where the contours allow a flux estimate at $\SI{144}{MHz}$ of the order of $\sim\SI{10}{mJy}$, which would lead to a spectral index of less than $-1$. This might indicate an aged radio source.	
	A precise measurement of the spectral index of this source would reveal more about its origin.

	Apart from the region G2 and G3 pointed out in Fig. \ref{fig:galaxy_density}, we can see more galaxy overdensities outside the clusters A\,1758\,N/S, which supports the idea of the two clusters being part of a large scale filament. Unfortunately, due to the complexity of the system, it is hard to model and look for residuals in the X-ray images (as demonstrated in the easier case of the region between the two clusters). The fact that we see many weaker edges in the X-ray surface brightness (see Tab. \ref{tab:sbredge}) may suggest the presence of weaker shocks from infalling filamentary material.
	
	\subsection{Galaxy structure and classification}
	We demonstrated in Section \ref{ch:galaxies} that a classification of the member galaxies based only on the sky location is inconclusive because of 1) the arbitrary definition of the thresholds, and 2) because the Gaussian components of the total distribution does not match the single Gaussians fit to spatially separated subsamples (e.g., N/S) of galaxies (see Fig. \ref{fig:redshift_histogram}).
	Using a Gaussian Mixture Model we identified in the three-dimensional phase space the classes of galaxies belonging to either south, north or an envelope surrounding both clusters with a peaked velocity distribution, shown in blue in Fig. \ref{fig:GMM_ALL}. We also noticed that spatial distribution of the newly classified northern and southern galaxies represents better the assumed merger axis (south-east to north-west for the northern cluster, north-east to south-west for the southern cluster). 
	When we used the same algorithm to subclassify the galaxies of the northern and southern cluster (Fig. \ref{fig:GMM_N} and \ref{fig:GMM_S}) we can identify the likely constituents of the merger in the north and south, and we also saw in the north a population of galaxies moving almost entirely in the plane of the sky (light blue in Fig. \ref{fig:GMM_N}). 
	The western galaxies in the northern cluster (red in Fig. \ref{fig:GMM_N}, ``West'' in Tab. \ref{tab:gmm_results}) show a relatively narrow distribution in velocity, and include all galaxies from the infalling group of galaxies detected first in the X-ray by \cite{Haines2018-re} . From their velocity dispersion we estimated roughly a virial mass for the group of $\SI{2.6(7)e13}{M_\odot}$. This is about 35\% lower than the mass estimated from the X-ray luminosity of that group ($\SI{4.1(8)e13}{M_\odot}$, \citealp{Haines2018-re}). The difference may be due to an excess in X-ray luminosity caused by the infall into the ICM of A\,1758\,N.
	
	\section{Summary and Conclusions}\label{ch:conclusion}
	The system Abell\,1758, consisting of two clusters, Abell\,1758\,N and S, will eventually merge to become an exceptionally massive cluster. Each of the two clusters is itself a merger, making the whole system a quadruple merger. This complex system offers the opportunity to study many merger-related phenomena, such as the impact of the merger on the gas in clusters, occurrence of shocks and cold fronts, radio halos and relics, and the interaction between clusters at a very early stage of the merging process (N and S). 
	We can now, better than ever before, characterize the mergers in the Abell\,1758 system:
	\begin{itemize}
	    \item For the first time we detect a shock front in Abell\,1758\,N. The Mach number ($\mathcal{M}\approx1.6$) is consistent when derived from the density and temperature discontinuity. This implies a relative merger velocity of about $\SI{2100}{km\,s^{-1}}$. Since the merger is roughly in the plane of the sky (as indicated by the X-ray gas distribution and the velocity information of the member galaxies), we can derive a timescale after the merger event of $\sim\SIrange{300}{400}{Myr}$, which is in excellent agreement with simulations. The shock front may be more extended than the region we analyze here (see Region 2 in Fig. \ref{fig:rggm}).
	    \item Using a 3D Gaussian Mixture Model approach we find that the galaxy population of the whole system can be subdivided into north (N) and south (S) components, with field galaxies in a very narrow velocity distribution around the two clusters. Subdivisions of the N and S populations show broader distributions (spatially and in the velocity domain), and a population confined in velocity and located between the merging cores of A\,1758\,N. While we cannot separate the galaxies of the NE and NW subclusters, we are able to identify galaxies scattered by core passage with a large velocity difference.
	    \item We find a hot envelope in the ICM of the northern cluster at the distance of the shock front. Although the merger in this cluster is past the core passage phase, the cool cores of the two subclusters can still be identified, and are connected by a region of cooler gas, likely stripped from the core regions.
	    \item The effects of the merger on the distribution of heavy elements is visible especially in the northern cluster: We find several peaks in the abundance map outside the cores of the northern subclusters, two located in front of the cores (in the projected direction of motion), while one is located between the cores.
	\end{itemize}
	We provide new hydrostatic mass estimates (assuming an NFW model) which we believe to be the most accurate X-ray mass estimates for A\,1758\,N and S. They are in good agreement with literature values and indicate an equal mass merger in the north (NW and NE subclusters have $\num{e15}$ and $\SI{1.2e15}{M_\odot}$, respectively), while the southern cluster has a total mass similar to one of the subclusters of A\,1758\,N. Both clusters, A\,1758\,N and S, are highly disturbed, so even deeper X-ray observations likely will not overcome systematic uncertainties connected with the hydrostatic mass estimation.
	
	The question of whether both clusters are part of a larger filament cannot be clearly answered: The X-ray data is inconclusive since we do not find excess emission with respect to the halos of the two clusters in the region of the overlap. We also do not confirm previous findings of an equatorial shock located between the northern and southern cluster, originating from the merger of the northern subclusters.
	On the other hand the galaxy distribution in the three dimensional phase-space indicates significant overlap of the two clusters.
	Furthermore, we find several interesting groups of galaxies near the northern cluster, some of them even associated with X-ray and diffuse radio emission. One infalling group had already been discovered in the X-ray regime, and we are able to identify the associated galaxies, even outside the detected X-ray halo of the group. The origin of the radio emission associated with this particular infalling group is not clear and needs further high resolution radio observations at the lowest frequencies. 
	
	It is also still unclear why there is no radio relic detected in the northern cluster. It might be blended with the extended radio halo. The discovered relic in the south is only found at low significance, and needs to be understood in more detail.
	
	\section*{Acknowledgements}
	The authors would like to thank Simona Giacintucci for helpful discussions.
	Support for this work was provided by the National Aeronautics and Space Administration
	(NASA) through \textit{Chandra} Award Numbers GO4-15129X, GO5-16135X, and GO5-16137X issued
	by the \textit{Chandra} X-ray Observatory Center (CXC), which is
	operated by the Smithsonian Astrophysical Observatory (SAO)
	for and on behalf of NASA under contract NAS8--03060.
	This research made use of NASA's Astrophysics Data System Bibliographic Services, and Astropy, a community-developed core Python package for Astronomy.  
	The authors thank the staff of the GMRT that made these observations possible. GMRT is run by the National Centre for Radio Astrophysics of the Tata Institute of Fundamental Research.
	We acknowledge usage of the NASA/IPAC Extragalactic Database (NED), which is operated by the Jet Propulsion Laboratory, California Institute of Technology, under contract with the NASA.
	
	\bibliographystyle{aasjournal}
	\bibliography{AstroAstro}
		
	\appendix
	\section{Plots}
	
	\begin{figure}[b]
		\includegraphics[width=0.45\textwidth]{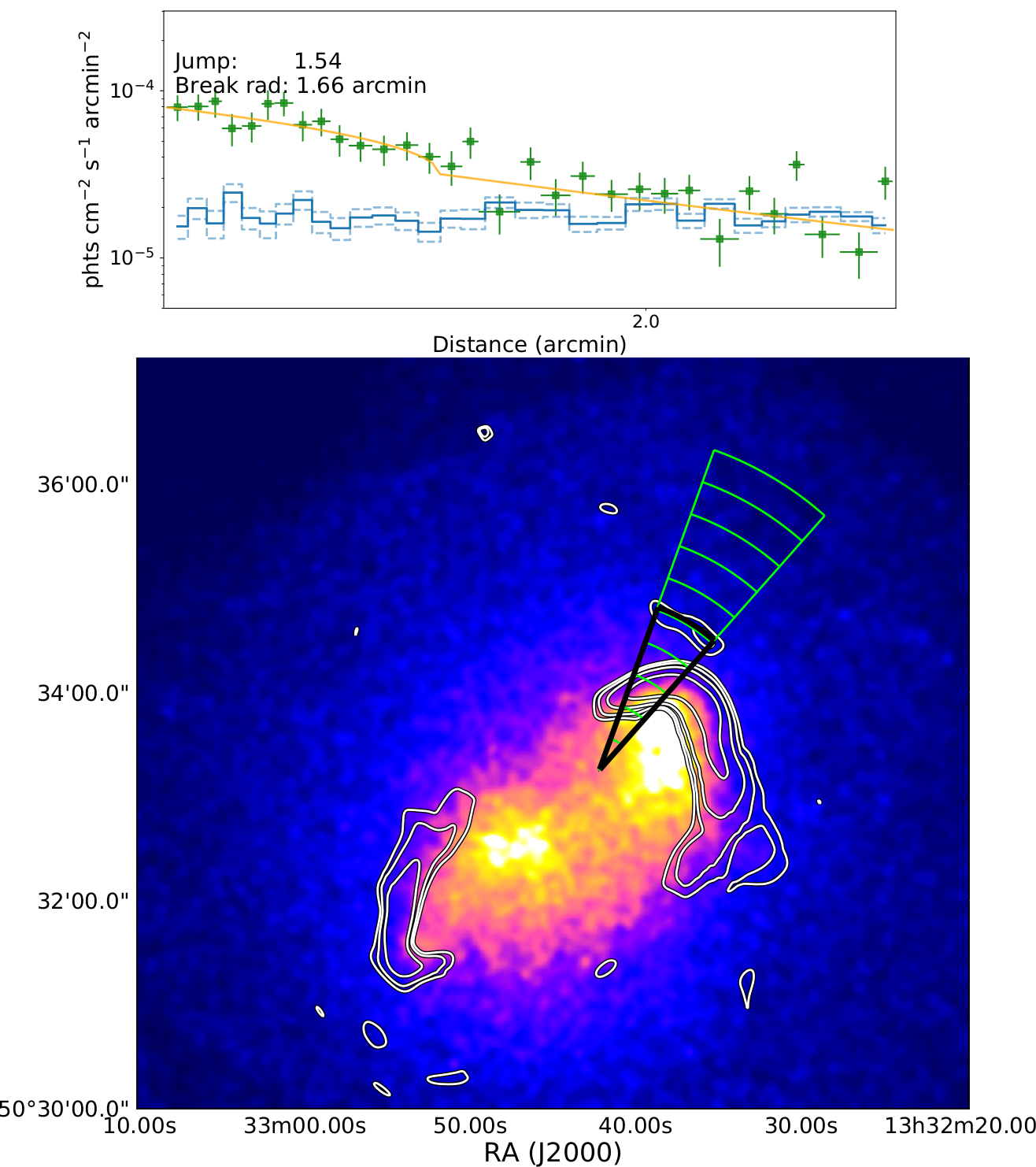}
		\includegraphics[width=0.45\textwidth]{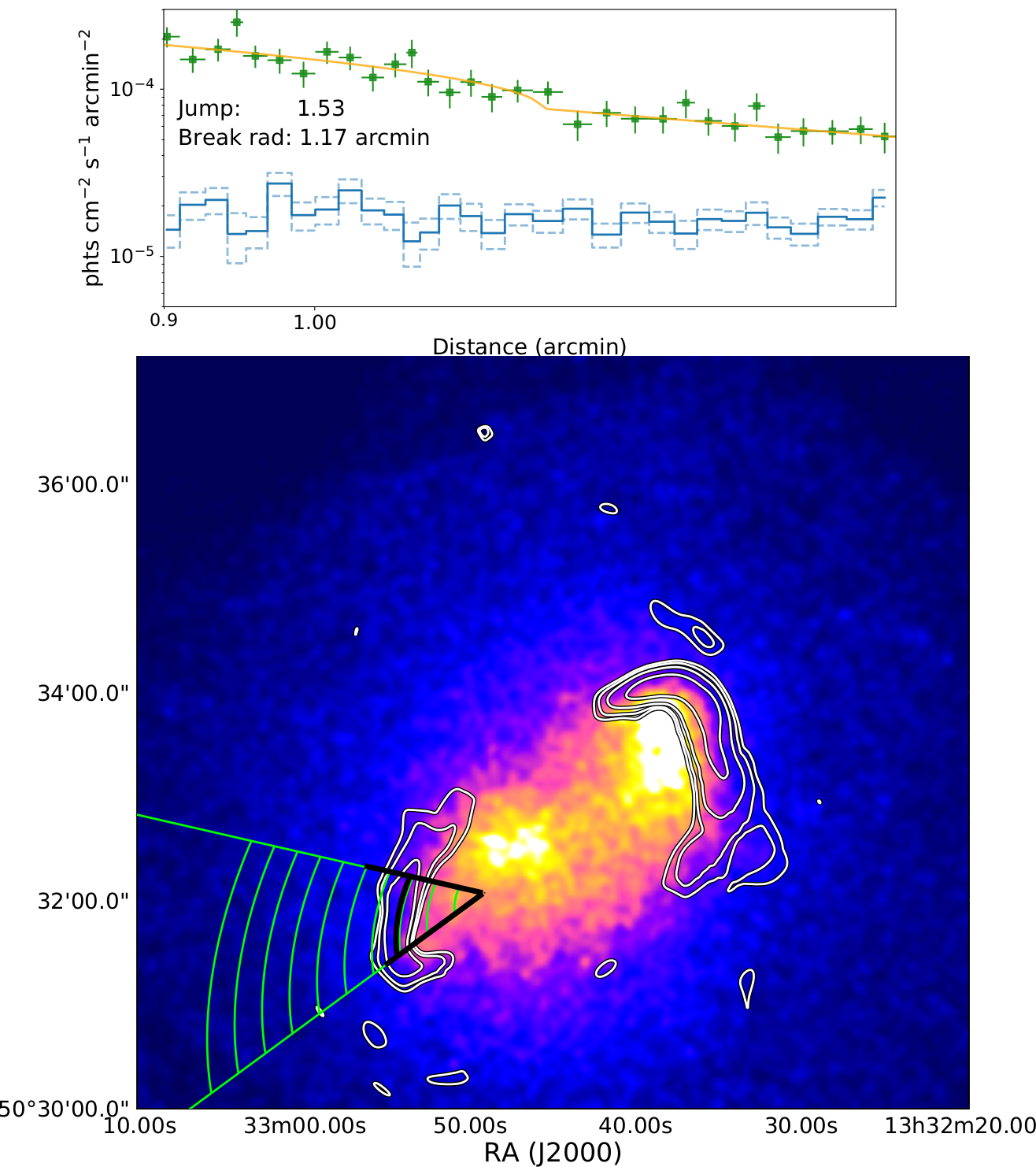}\vspace{10px}\\
		\includegraphics[width=0.45\textwidth]{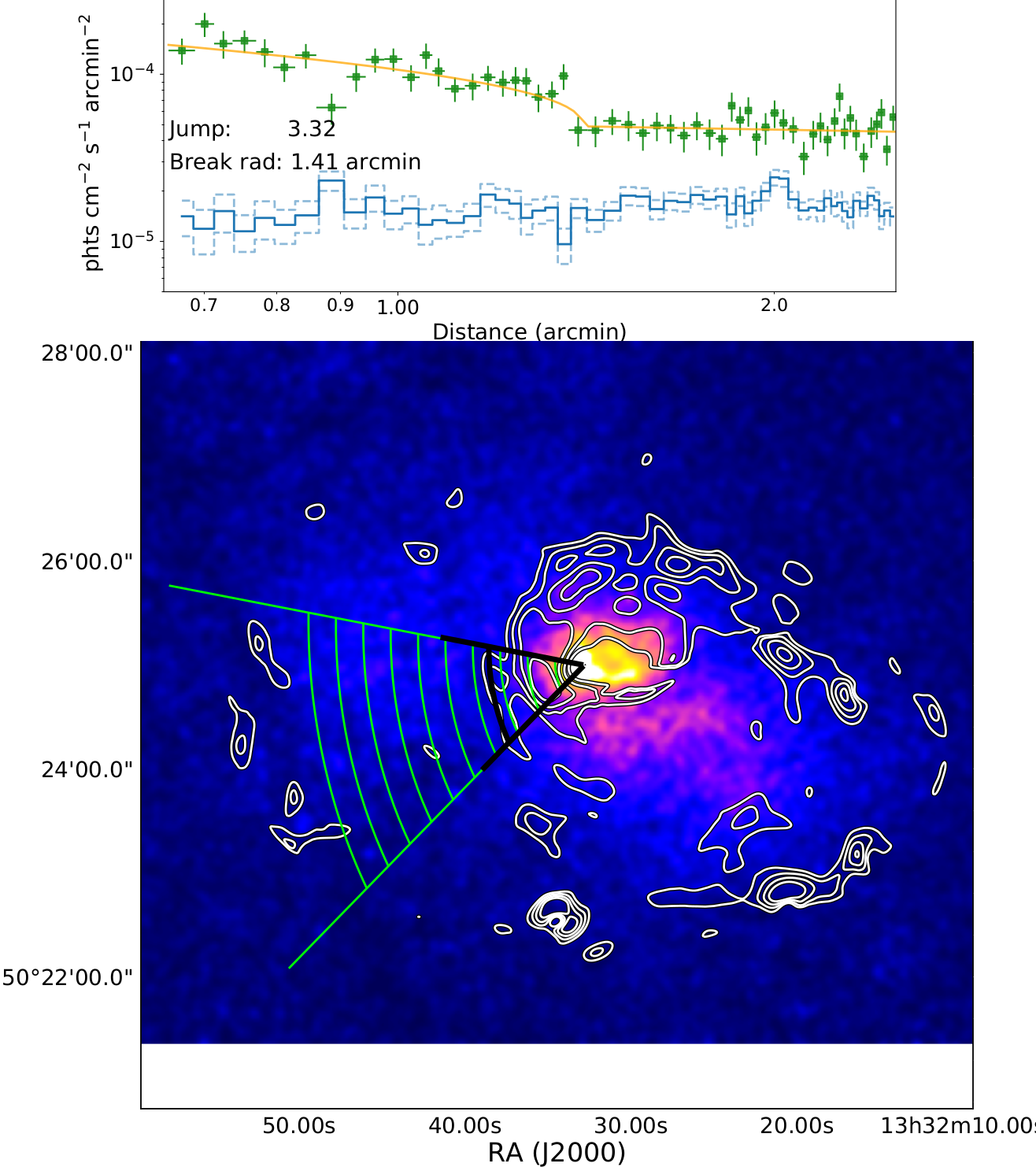}
		\includegraphics[width=0.45\textwidth]{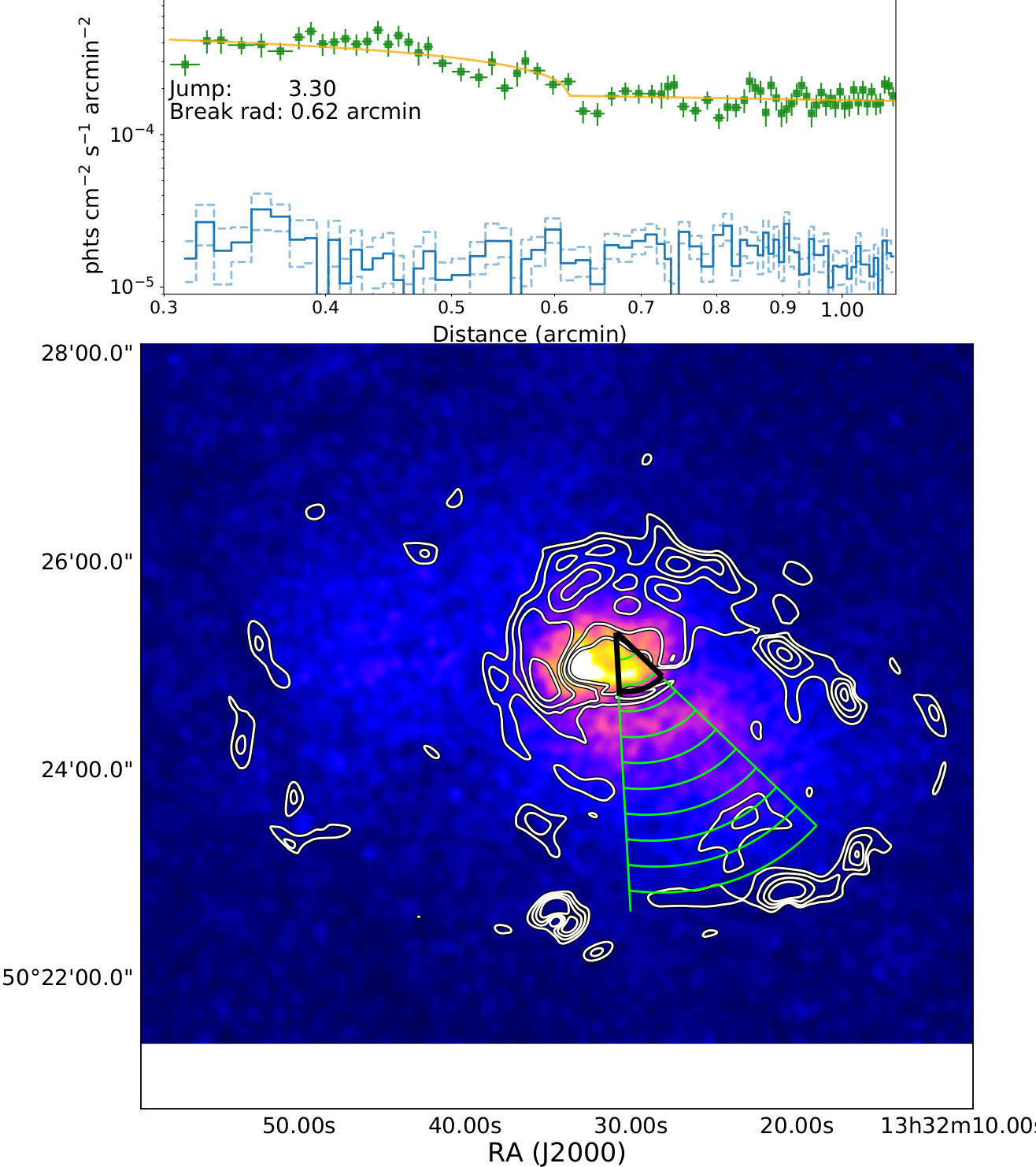}
		\caption{Each of the four panels (top left: region 2, top right: region 3, bottom left: region 4, bottom right: region 5, regions are labeled according to Tab. \ref{tab:sbredge}) shows the one-dimensional surface brightness distribution (top plot in each panel, background in blue, background subtracted data in green, model in yellow), along the region indicated in the bottom smoothed \textit{Chandra} image in green (bottom plot in each panel, contours show the edges seen in the gradient image, and the thick black line shows the location of the discontinuity). }
		\label{fig:region2}
	\end{figure}
		
	\begin{figure}		
		\includegraphics[width=0.45\textwidth]{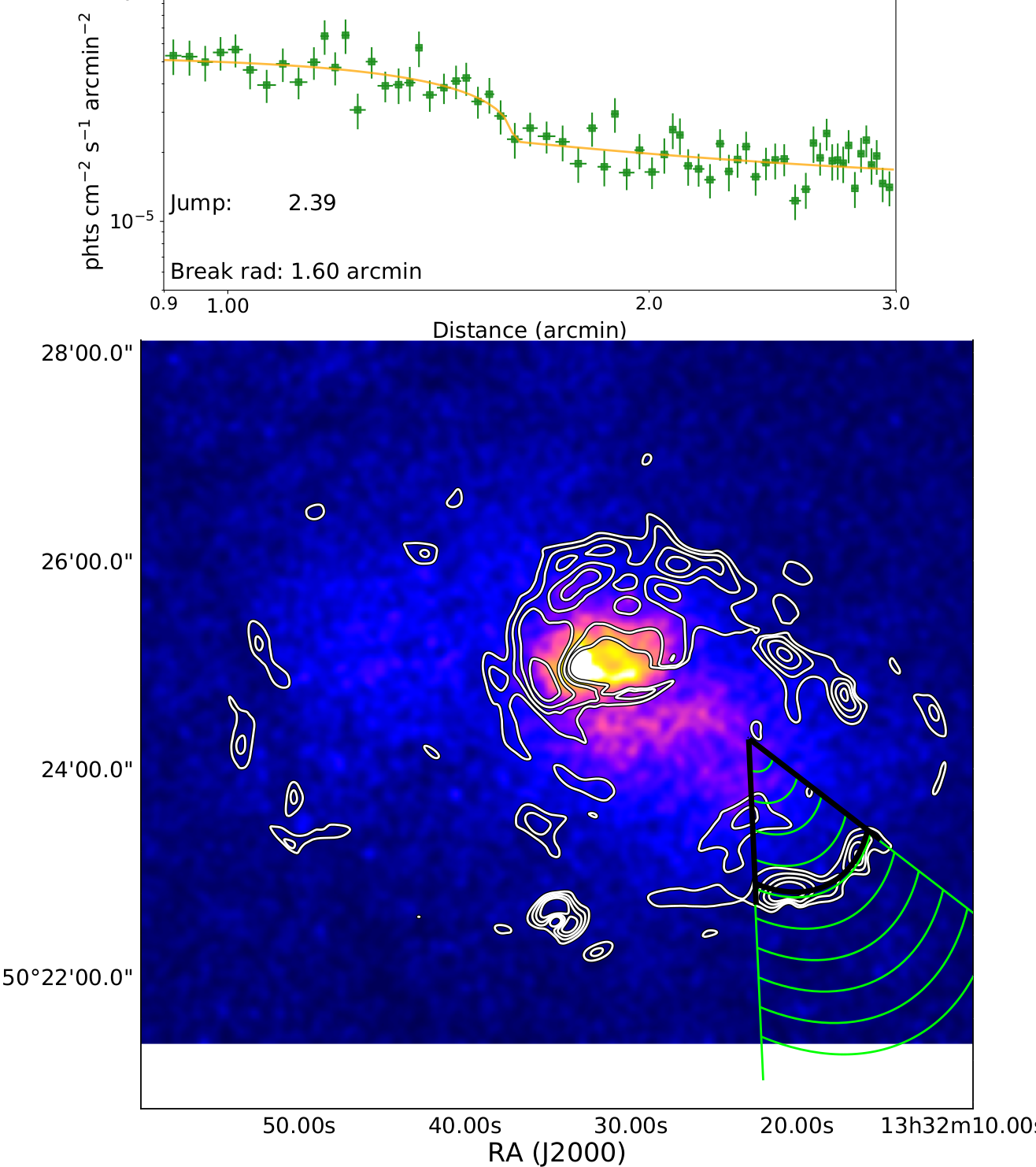}
		\includegraphics[width=0.45\textwidth]{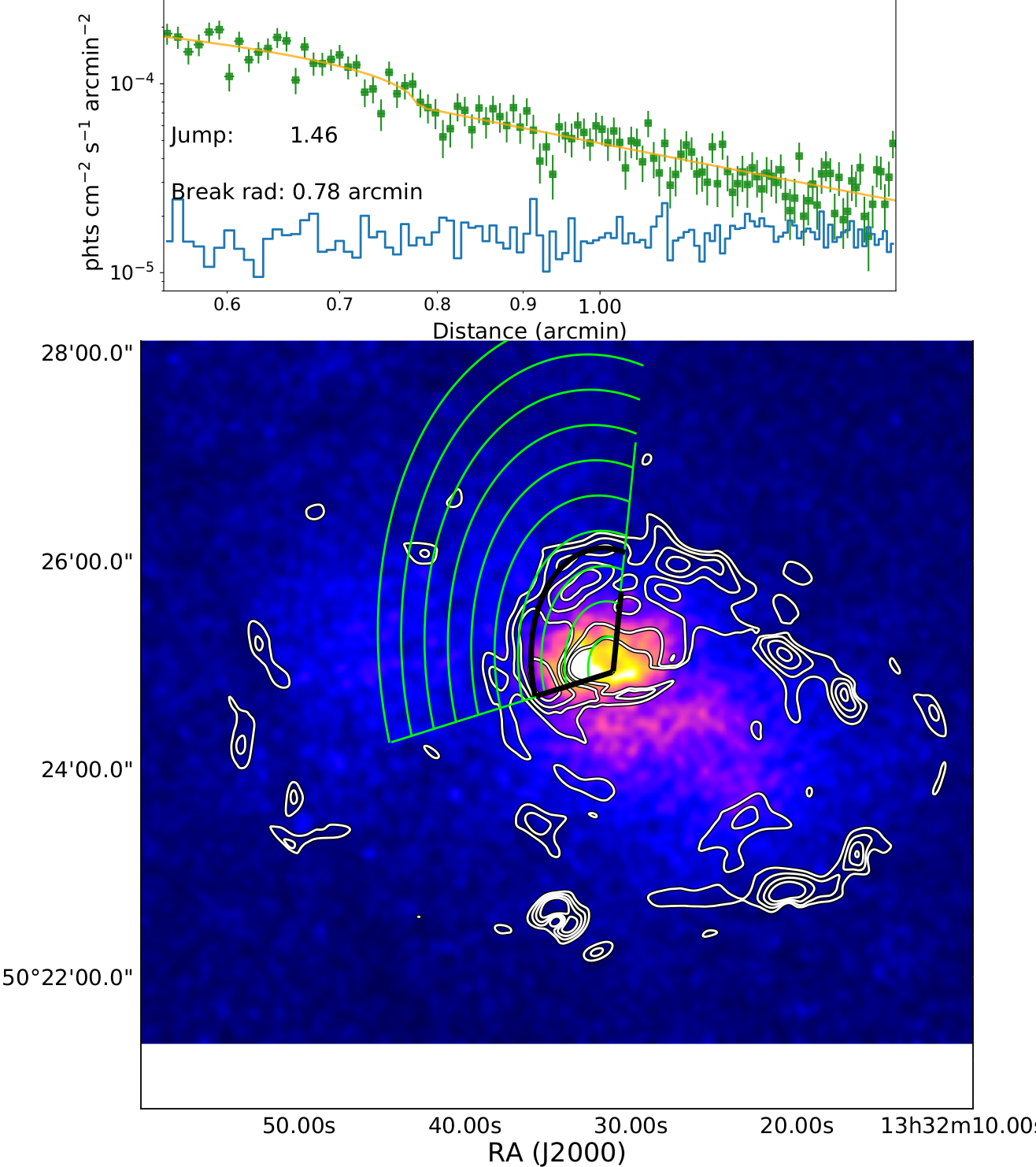}
		\caption{As Fig. \ref{fig:region2} but only with two panels for region 6 (left) and 7 (right).}
		\label{fig:region3}
	\end{figure}
	
\end{document}